\documentclass[prc,amsmath,twocolumn]{revtex4} 

\usepackage{dcolumn}
\usepackage{graphicx}
\usepackage{color}

\newcommand*{\s}[1]{/\llap{$#1$}} 
\newcommand*{\lrpart}{\tensor\partial}
\newcommand*{\dop}[2]{#1\cdot#2}
\newcommand*{\br}[1]{\mathcal{A}_\chi(#1)}
\newcommand*{\brs}[1]{\mathcal{A}_\chi(\s#1)}

\newcommand*{\gbrs}[1]{\gamF\mathcal{A}_\chi(\s#1)}
\newcommand*{\res}[2]{\ensuremath{#1_{#2}}}
\newcommand*{\rres}[3]{\ensuremath{#1_{#2}(#3)}}
\newcommand*{\gamF}{\gamma_5}
\newcommand*{\kapp}{\kappa_p}

\newcommand*{\eps}{\varepsilon}
\newcommand*{\mk}{\ensuremath{{m_K}}}

\newcommand*{\A}{\ensuremath{\mathsf A}}
\newcommand*{\M}{\ensuremath{\mathsf M}}
\newcommand*{\E}{\ensuremath{\mathsf E}}
\newcommand*{\K}{\ensuremath{\mathsf K}}
\newcommand*{\T}{\ensuremath{\mathsf T}}
\newcommand*{\tf}{\ensuremath{\tilde f}}
\newcommand*{\tF}{\ensuremath{\mathsf G}}
\newcommand*{\La}{\ensuremath{\mathcal L}}

\newcommand*{\up}{u_p}
\newcommand*{\us}{\bar u_\Sigma}
\newcommand*{\e}{\ensuremath{e_K}}
\newcommand*{\oh}{\ensuremath{\frac12}}
\renewcommand*{\th}{\ensuremath{\frac32}}

\newcommand*{\mc}[1]{\multicolumn{1}{c|}{#1}}
\newcommand*{\mcc}[1]{\multicolumn{1}{c}{#1}}
\newcommand*{\N}{\mc{---}}  
\newcommand*{\Nc}{\mcc{---}}  
\newcommand*{\X}[1]{\begin{cases} #1 \end{cases}}
\newcommand*{\ch}[4]{\ensuremath{#1 #2 \to #3 #4}}
\long\def\Omit#1{}

\DeclareMathOperator{\diag}{diag}
\DeclareMathOperator{\Tr}{Tr}

\newcolumntype{x}[1]{D..{#1}}
\newcolumntype{C}{>{$}c<{$}}
\newcolumntype{R}{>{$}r<{$}}

\newcommand*{\tblref}[1]{Table~\ref{tbl:#1}}
\newcommand*{\tbllab}[1]{\label{tbl:#1}}
\renewcommand*{\eqref}[1]{Eq.~(\ref{eq:#1})}
\newcommand*{\eqlab}[1]{\label{eq:#1}}
\newcommand*{\figref}[1]{Fig.~\ref{fig:#1}}
\newcommand*{\figlab}[1]{\label{fig:#1}}
\newcommand*{\secref}[1]{Section~\ref{sec:#1}}
\newcommand*{\seclab}[1]{\label{sec:#1}}

\def\VYP#1#2#3{{\bf #1}, #3 (#2)}  

\def\NPA#1#2#3{Nucl.~Phys.~A~\VYP{#1}{#2}{#3}}

\def\PLB#1#2#3{Phys.~Lett.~B~\VYP{#1}{#2}{#3}}

\def\PRC#1#2#3{Phys.~Rev.~C~\VYP{#1}{#2}{#3}}
\def\PRD#1#2#3{Phys.~Rev.~D~\VYP{#1}{#2}{#3}}
\def\PRL#1#2#3{Phys.~Rev.~Lett.~\VYP{#1}{#2}{#3}}

\def\EPJA#1#2#3{Eur.\ Phys.\  J.\ A\ \VYP{#1}{#2}{#3}}
\newcommand{\m}[1]{#1}

\begin{document}

\title{$K\Lambda$ and $K\Sigma$ photoproduction in a coupled
channels framework.}

\author{A. Usov}
\email{usov@kvi.nl}
\author{O. Scholten}
\email{scholten@kvi.nl}
\affiliation{Kernfysisch Versneller Instituut, University of Groningen,
9747 AA, Groningen, The Netherlands}

\begin{abstract}
A coupled channels analysis, based on the \K-matrix approach, is presented for
photo-induced kaon production. It is shown that channel coupling effects are
large and should not be ignored. The importance of contact terms in the
analysis, associated with short range correlations, is pointed out. The
extracted parameters are compared with $SU(3)$-model predictions.
\end{abstract}
\maketitle

\section{Introduction}

One of the major issues of hadron physics is the determination of
masses and coupling constants for the different baryon resonances.
These extracted parameters serve as a test of different QCD-based
models~\cite{Met01,The01}. As will be
demonstrated in this work, coupled channels effects are large and
should be taken into account in extracting resonance parameters
for the higher-lying resonances.
Another reason for performing a coupled channels
description is that the requirement of a simultaneous fit of the
data for a multitude of observables in different reaction channels
strongly confines the values of the model parameters thus reducing
the model dependence to a minimum. The coupled-channel calculations
presented here are based on an effective-field model which
is gauge invariant and obeys the low-energy theorems.

However, as it was shown for example in ref.~\cite{Pen02}, even the present
large experimental database in a unitary coupled-channel effective Lagrangian
model does not allow to uniquely fix the extracted parameters. One of the
reasons for this is the necessity to include empirical form-factors in the
model to regularize the matrix elements at higher energies. These form-factors
introduce the need for a gauge-invariance restoration procedure. As is well
known this has many ambiguities associated with it, see for example
ref.~\cite{Dav01}. In this work we will explicitly formulate these ambiguities
in terms of four-point contact terms which can be added to the model
Lagrangian. In particular we will show that procedure of minimal substitution,
known as the Ohta prescription~\cite{Ohta89} results in a major cancellation of
the form-factor effects. This results in a large disagreement with the data,
even in a coupled-channels description, which has the tendency to suppress the
cross section at higher energies in a particular channel.

A number of analyses of data on strangeness production have been
performed~\cite{Tit02,Ire04,Jan02,Pen02,Benn01}, but only few of them are
based on a coupled-channel model~\cite{Pen02}. In addition different
gauge-invariance restoration prescriptions are used, which makes it difficult
to compare the parameters. In the present work we have investigated the
implications of the different gauge-invariance restoration procedures,
and formulated them in terms of
additional gauge-invariant contact terms. In general these contact terms can be
regarded as short-range effects which are not included in the model Lagrangian
or due to loop corrections which have been omitted. In expressing the model
dependence in terms of contact terms added to the model Lagrangian we follow
the philosophy used in effective-field chiral perturbation theory~\cite{Mei95}.

In the last years the data base on strangeness photoproduction~\cite{SAPHIR1,
SAPHIR2} has been extended appreciably. For certain kinematics these new data
differs significantly from the old one. A new analysis is therefore appropriate.

\section{Model}

This work is based on an effective Lagrangian model. The \m{Lagrangian as used
in the present calculation} is given in the Appendix~\ref{app:lagrangian} and
some of the main ingredients are presented in a following section. This
Lagrangian is used to build the kernel for a \K-matrix approach. As described
in the following section this allows to account for coupled channels effects
while preserving many symmetries of a full field-theoretical approach.

\subsection{\K-matrix model}\label{sec:Kmatr}

In our calculation the coupled channels (or re-scattering) effects are included
through the use of the \K-matrix formalism. In this section we present a short
overview of the \K-matrix approach, a more detailed description can be found in
ref.~\cite{Kor98,Sch02,New82}.

In the \K-matrix formalism the scattering matrix is written as
\begin{equation}\eqlab{T-matr}
  \T = \frac{\K}{1-i\K} \,.
\end{equation}
It is easy to check, that the resulting scattering amplitude $S=1+2i\T$
is unitary provided that \K\ is Hermitian. The construction in
\eqref{T-matr} can be regarded as the re-summation of an infinite
series of loop diagrams by making a series expansion,
\begin{equation}
  \T = \K + i\K\K + i^2\K\K\K + \cdots \,.
\end{equation}
The product of two \K-matrices can be rewritten as a sum of different one-loop
contributions (three- and four-point vertex and self-energy corrections)
depending on the Feynman diagrams which are included in the kernel, \K.
However, not the full spectrum of loop corrections present in a true
field-theoretical approach, is generated in this way and the missing ones
should be accounted for in the kernel. In constructing the kernel one should be
careful to prevent double counting. For this reason we include in the kernel
tree-level diagrams only, modified with form-factors and contact terms. The
contact terms (or four-point vertices) ensure gauge invariance of the model and
express model-dependence in working with form factors, see~\secref{ff}. Form
factors and contact terms can be regarded as accounting for loop corrections
which are not generated in the \K-matrix procedure, or for short-range
effects which have been omitted from the interaction Lagrangian. Both s- and
u-channel diagrams are included and the kernel thus obeys crossing symmetry.

{\squeezetable
\begin{table}
  \caption{ \tbllab{resonances}
    Baryon states included in the calculation of the
    kernel with their coupling constants. The column labelled
    W lists the decay width to states outside the model space.
    The columns labelled M and W are in units of GeV. See text
    for a discussion on the signs of the coupling constants.}
  \begin{ruledtabular}
  \begin{tabular}{C|x3|x3|d|d|d|d|d|d} 
    L_{IJ} &\mc{M}&\mc{W}&\mc{$g_{N\pi}$}
            &\mc{$g^1_{N\gamma}$}&\mc{$g^2_{N\gamma}$}
             &\mc{$g_{K\Lambda}$}&\mc{$g_{K\Sigma}$}&\mcc{$g_{N\eta}$} \\
    \hline
    N           &0.939 &0.0  &13.47&\N & \N & 12  &8.7 &3.0 \\
    \Lambda     &1.116 &0.0  &\N & \N  & \N & \N  &\N  & \Nc\\
    \Sigma      &1.189 &0.0  &\N & \N  & \N & \N  &\N  & \Nc\\
    \hline
    \res{S}{11} &1.525 &0.0  &0.6&-1.2 & \N & 0.1 &0.0 &2.0 \\
    \res{S}{11} &1.690 &0.030&1.0&-1.0 & \N &-0.1 &0.0 &-0.5\\
    \res{S}{31} &1.630 &0.100&3.7&-0.25& \N & \N  &-0.8& \Nc\\
    \hline
    \res{P}{11} &1.480 &0.200&5.5& 1.0 & \N &0.0  &-2.0&0.0 \\
    \res{P}{11} &1.750 &0.300&3.0& 0.3 & \N &0.0  &-6.0&0.0 \\
    \res{P}{13} &1.750 &0.300&0.12&-0.5& 2.0&-0.035&0.0&0.0 \\
    \res{P}{33} &1.230 &0.0  &1.7&-2.2 &-2.7& \N  &0.0 & \Nc\\
    \res{P}{33} &1.855 &0.150&0.0&-0.4 &-0.6& \N  &0.55& \Nc\\
    \hline
    \res{D}{13} &1.515 &0.050&1.2& 5.0 & 5.5&2.0  &0.0 &2.0 \\
    \res{D}{13} &1.700 &0.090&0.0&-1.0 & 0.0&0.0  &0.3 &0.0 \\
    \res{D}{33} &1.670 &0.250&0.8& 1.5 & 0.6& \N  &-3.0& \Nc\\
  \end{tabular}
  \end{ruledtabular}
\end{table}
}

\begin{table}
  \caption{ \tbllab{particles}
    Mass, spin, parity and isospin of the mesons which are
    included in the model. The rightmost column 
    specifies in which reaction channels their t-channel
    contribution are taken into account. }
  \begin{ruledtabular}
  \begin{tabular}{C|d|CC|c}
    \mc{Meson}&\mc{M [GeV]}& S^\pi & I & t-ch contributions \\
    \hline
    \pi     & 0.135     & 0^- & 1 & (\ch \gamma N \phi N), (\ch \pi N \rho N)\\
    K       & 0.494     & 0^- &\oh& (\ch \gamma N K\Lambda), (\ch \gamma N K\Sigma) \\
    \phi    & 1.019     & 1^- & 0 & \\
    \eta    & 0.547     & 0^- & 0 & (\ch \gamma N \phi N) \\
    \hline
    \rho    & 0.770     & 1^- & 1 & (\ch \gamma N \pi N), (\ch K\Lambda K\Sigma), \\
            &           &     &   & (\ch K\Sigma K\Sigma), (\ch N\pi K\Lambda), \\
            &           &     &   & (\ch N\pi N\eta), (\ch N\pi N\pi) \\
    \omega  & 0.781     & 1^- & 0 & (\ch N\gamma N\pi) \\
    \sigma  & 0.760     & 0^+ & 0 & (\ch N\gamma N\phi), (\ch N\pi N\pi) \\
    K^*     & 0.892     & 1^- &\oh& (\ch N\gamma K\Lambda),(\ch N\gamma K\Sigma), \\
            &           &     &   & (\ch K\Lambda N\eta), (\ch K\Sigma N\eta), \\
            &           &     &   & (\ch N\pi K\Sigma) \\
  \end{tabular}
  \end{ruledtabular}
\end{table}

To be more specific, the loop corrections generated in the \K-matrix procedure
include only diagrams that correspond to two on-mass-shell particles in the
loop~\cite{Kon00,Kon02}. This is the minimal set of diagrams one has to include
to ensure two-particle unitarity. Not included are thus all diagrams that are
not 2 particle reducible. In addition only the convergent pole contributions
i.e.\ the imaginary parts of the loop correction, are generated. \m{The omitted
real parts are important to guarantee analyticity of the amplitude and may have
complicated cusp-like structures at energies where other reaction channels
open. In principle these can be included as form factors as is done in the
dressed-\K-matrix procedure~\cite{Kon00,Kor03}.} For reasons of simplicity we
have chosen to work with purely phenomenological form factors in the present
calculations. \m{An alternative procedure to account for the real-loop
corrections is offered by the approach of~\cite{Sat96} which is based on the
use of a Bethe-Salpeter equation. This approach was recently extended to kaon
production in~\cite{Chi04}. Another possible approach is the one discussed
in~\cite{Lut02} which is based on a different application of the K-matrix
formalism.}

The strength of the \K-matrix procedure is that, in spite of its
simplicity, several symmetries are obeyed~\cite{Sch02}. As was
already noted the resulting amplitude is unitary, provided that
\K\ is Hermitian, and obeys gauge invariance since the kernel
is gauge-invariant. In addition the scattering amplitude obeys
crossing symmetry when the kernel is crossing symmetric. This
property is crucial for a proper behavior in the low-energy
limit~\cite{Kon02,Kon04} of the scattering amplitude.

Coupled-channels effects are automatically accounted for by the \K-matrix
approach for the channels explicitly included into the \K-matrix as the final
states. To account for the coupling to other channels we have added an explicit
dissipative part to the kernel.

The resonances which are taken into account in building the kernel are
summarized in \tblref{resonances}. \m{In the current work we limit ourselves to
the spin-$\frac12$ and $\frac32$ resonances as in this energy regime
higher-spin resonances are known~\cite{Shk04} to give only a minor contribution
to the strange channels, which are of primary interest here.}
\m{Spin 3/2 resonances are included with
so-called gauge-invariant vertices which have the property that the coupling to
the spin-1/2 pieces in the Rarita-Schwinger propagator
vanish~\cite{Pas00,Kon00}. We have chosen for this prescription since it
reduces the number of parameters as we do not have to deal with the
off-shell couplings. The effects of these off-shell couplings can be absorbed
in contact terms~\cite{Pas01} which we prefer, certainly within the context of
the present work.}
\m{
The masses of the resonances given in \tblref{resonances} are bare masses and
they thus may deviate from the values given by Particle Data Group~\cite{PDG}.
Higher-order effects in the K-matrix formalism do give rise to a (small)
shift of the pole-position with respect to the bare masses. The masses of very
broad resonances, in particular the $P_{11}$, are not well determined, a rather
broad range of values (typically a spread of the order of a quarter of the
width) gives comparable results.  The width quoted in
\tblref{resonances} corresponds to the partial width for decay to states outside
our model space.} The t-channel contributions which are included in the kernel
are summarized in \tblref{particles}.

\m{In the present calculation we have chosen all primary
coupling constants to the nucleon positive. In particular the sign of $g_{NK
\Lambda}$, see \tblref{resonances}, deviates from the customary negative
value~\cite{Cot04}. In a calculation like ours and many of the ones cited
in~\cite{Cot04} this sign is undetermined. Changing the sign of all coupling
constants involving a single $\Lambda$-field leaves the calculated observables
invariant since it corresponds to a sign redefinition of this field. In weak
decay the ratio of the vector v.s.\ axial-vector coupling does correspond to an
observable. The magnitude of the couplings is within the broad range specified
in~\cite{Cot04}.}

\subsection{Model space, channels included}

To keep the model manageable and relatively simple we consider only stable
particles or narrow resonances in 2-body final states which are
important for strangeness photoproduction. The
$K-\Lambda$, $K-\Sigma$, $\phi-N$, $\eta-N$ and $\gamma-N$ are the final states
of primary interest, and the $\pi-N$ final state is included for its strong
coupling to most of the resonances. Three-body final states, such as $2\pi-N$,
are not included explicitly for reasons of simplicity. Their influence on the
width of resonances in taken into account by assigning an additional (energy dependent)
width to resonances~\cite{Kor98}. To investigate the effects of the coupling to
more complicated states, we also included the $\rho-N$ final state. As
discussed in the results section, including the $\rho$ channel has a strong
influence on the pion sector, but has a relatively minor effect on $\Lambda$
and $\Sigma$ photoproduction, which are our primary focus. The discussion
of $\phi$-meson production will be presented in a subsequent paper.

\m{The components of the
kernel which couple the different non-electromagnetic channels are taken as the
sum of tree-level diagrams, similar to what is used for the photon channels.
For these other channels no additional parameters were introduced and they thus
need no further discussion.}

\section{Form-factors \& gauge restoration \seclab{ff}}

A calculation with Born contributions, without the introduction of
form factors, strongly overestimates the cross section at higher
energies. Inclusion of coupled-channels effects reduces the cross
section at high energies, however not sufficiently to obtain agreement
with the experimental data, and one is forced to quench the Born
contribution with form factors. There are two physical motivations for
introducing form factors (or vertex functions) in our calculation.
First of all, at high photon energies one may expect to become
sensitive to the short-range quark structure of the nucleon. Because
this is not included explicitly in our model we can only account for
it through the introduction of phenomenological vertex functions. A
second reason are intermediate-range effects due to meson-loop
corrections which are not generated through the \K-matrix formalism.
Examples of these are given in refs~\cite{Kon00,Kor03}.

In the approach followed in this paper and, for example, that of
ref.~\cite{Pen02}, the form-factors are not known \emph{a priori}
and thus they introduce a certain arbitrariness in the model. In the
current paper we limit ourselves to dipole form-factors in the
s-, t-, and u-channels because of their simplicity,
\begin{equation}\eqlab{ff-dipole}
  F_m(s)=\frac{\Lambda^2}{\Lambda^2+(s-m^2)^2} \;.
\end{equation}
For ease of notation we introduce the subtracted form factors
\begin{equation}\eqlab{ff-twiddle}
  \tf_m(s)=\frac{1-F_m(s)}{s-m^2} \,,
\end{equation}
where $F_m(s)$ is normalized to unity on the mass-shell, $F_m(m^2)=1$, and
$\tf_m(m^2)$ is finite. In the kaon sector only we use a different functional
form for the u-channel form-factors
\begin{equation}\eqlab{ff-u-channel}
  H_m(u)=\frac{u\Lambda^2}{\big( \Lambda^2+(u-m^2)^2 \big)m^2} \;,
\end{equation}
where the argumentation for different choice is presented in the discussion
of the $\Sigma$ photoproduction results. Often a different functional
form and cut-off values are introduced for t-channel form-factors. While this
can easily be motivated, it introduces additional model dependence and
increases the number of free parameters. To limit the overall number of
parameters we have taken the same cut-off value ($\Lambda=1.2\; \mbox{GeV}^2$,
see \eqref{ff-dipole}) for all form-factors except for Born contributions in
kaon channels where we used $\Lambda=1.0\;\mbox{GeV}^2$.

Inclusion of form-factors will in general break electromagnetic
gauge-invariance of the model. A gauge-restoration procedure thus should be
applied. To see the implications of the gauge restoration procedure we take as
a specific example the $(\gamma+p \to K+\Sigma)$ amplitude. To keep our
discussion at the most general level we allow for the use of different
form-factors for the different contributions to the amplitude. In the
tree-level approximation, with form-factors included, the scattering
amplitude reads
\begin{multline}\eqlab{born}
  \dop{\eps}{M_{fi}} = g_{N\Sigma K} \us \Big[F_m(s) \gbrs{q}
   \frac{\s{p}+\s{k}+m}{s-m^2} \\
  \times (\s{\eps} - \frac{\kapp}{2m} \s{\eps}\s{k})
    +F_{m_\Sigma}(u)
      \bigl( (1-\e)\s{\eps} - \frac{\kappa_\Sigma}{2m}\s{\eps}\s{k} \bigr) \\
  \times \frac{\s p'-\s k+m_\Sigma}{u-m_\Sigma^2} \gbrs{q} \\
  +\dop\eps{(2q-k)} \frac{\e F_\mk(t)}{t-\mk^2} \gbrs{q-\s{k}} \Big]\up \;,
\end{multline}
where $\up$ and $\us$ are the nucleon and $\Sigma$ spinors; $p$, $k$, $q$
and $p'$ are the momenta of the incoming proton, photon and outgoing $K$
and $\Sigma$ respectively, and $\eps$ is the photon polarization vector.
The charge of the outgoing kaon is $\e$ and
\begin{equation}
  \brs{q}=\frac{\chi+\s q/2m}{\chi+1}
\end{equation}
is a short-hand notation introduced to account for both pseudo-vector
($\chi=0$) and pseudo-scalar ($\chi\rightarrow\infty$) couplings. While
all derivations are done for both types of couplings, the final calculation
is done using pseudo-vector couplings for both the pion and the kaon.
For simplicity we have omitted from \eqref{born} the contribution with
an intermediate $\Lambda$ since it is gauge invariant,
\begin{multline}
  \dop{\eps}{M_{fi\Lambda}} =
    -g_{N\Lambda K}\e F_{m_{\Lambda}}(u) \frac{\kappa_{\Lambda\Sigma}}{2m} \\
    \times \us \s{\eps}\s{k}
      \frac{\s{p}'-\s{k}+m_{\Lambda}}{u-m_{\Lambda}^2} \gbrs{q} \up \;.
\end{multline}
It is easy to check that \eqref{born} is not gauge invariant:
\begin{multline}
  \dop{k}{M_{fi}} = g_{N\Sigma K} \us \gamF \Big[
    \brs{q} \bigl( F_m(s) - (1-\e)F_{m_\Sigma}(u) \bigr) \\
    - \br{\s{q}-\s{k}} \e F_\mk(t)
    \Big]\up \neq 0  \,.
\end{multline}
This result was to be expected since form factors should be introduced at the
Lagrangian level and some kind of (minimal) substitution procedure should have
been followed to obtain the photon vertices necessary to formulate a
gauge-invariant theory.

\subsection{Gauge Restoration}

In this section different gauge-restoration procedures are compared using the
amplitude for the $(\gamma+p\to K+\Sigma)$ reaction as a specific example. The
procedure of minimal substitution will generate the necessary vertex
corrections and contact terms to restore gauge invariance. We will follow
rather closely the notation introduced in ref.~\cite{Kon00,Kon02}.
Thus minimal substitution in the strong vertex dressed with form factors,
gives
\begin{widetext}
\begin{multline}\eqlab{cnt-term-1}
  F_m(p^2) F_\mk(q^2) F_{m_\Sigma}(p'^2)\gbrs{q} \longmapsto
    (2p+k)^\mu \tf_m(s) \gbrs{q} \\
   +(1-\e)(2p'-k)^\mu \tf_{m_\Sigma}(u) \gbrs{q}
   +\e (2q-k)^\mu \tf_\mk(t) \gbrs{q}
   -\frac{\e F_\mk(t)}{2m(\chi+1)}\gamF \gamma^\mu \;,
\end{multline}
where we have dropped the coupling constant. We should note that this result is
far from unique and depends on the exact order of performing minimal
substitution~\cite{Kon00}. For example, if we take an ordering of the terms
which is more symmetric for the in- and out-going baryon, we obtain for the
case of pseudo-vector coupling for charged kaon production (\e=1)
\begin{subequations}\eqlab{cnt-term-2}
\begin{multline} \eqlab{cnt-term-2a}
  \gamF \s{q} F_\mk(q^2) F_m((q+p')^2)  \longmapsto
     (2p+k)^\mu \tf_m(s) F_\mk(t) \gamF (\s{q}-\s{k}) \\
    +(2q-k)^\mu \tf_\mk(t) F_m(s) \gamF (\s{q}-\s{k})
    -\frac{F_m(s)}{2m(\chi+1)}\gamF \gamma^\mu \,,
\end{multline}
and for neutral kaon production (\e=0)
\begin{multline} \eqlab{cnt-term-2b}
  \gamF\Big(
     \s{p} F_m(p^2) F_{m_\Sigma}((p-q)^2)
    -\s{p}' F_{m_\Sigma}(p'^2) F_m((p'+q)^2) \Big) \longmapsto
     \frac{F_{m_\Sigma}(u)-F_m(s)}{2m(\chi+1)}\gamF \gamma^\mu \\
    +\Big( (2p+k)^\mu \tf_m(s) F_{m_\Sigma}(u) + (2p'-k)^\mu
            \tf_{m_\Sigma}(u) F_m(s) \Big)
  \gamF (\s{q}-\s{k}) \,,
\end{multline}
\end{subequations}
\end{widetext}
which substantially differs from \eqref{cnt-term-1}. It can easily be
checked that both \eqref{cnt-term-1} and \eqref{cnt-term-2} in
combination with the born contribution \eqref{born} give rise to a
gauge-invariant expressions.

The effects of the gauge-restoration procedure are more easily seen when one
rewrites the amplitude in terms of gauge-invariant amplitudes
$\dop\eps{M'_{fi}} = \sum_i \us \M_i\A_i \up$, where gauge-invariant operators
are given as
\begin{equation}
  \begin{split}
    \M_1&=-\gamF\s\eps\s k \\
    \M_2&=2\gamF(\dop p\eps \dop{p'}k-\dop{p'}\eps \dop p k)\\
    \M_3&=\gamF(\dop p k\s\eps-\dop p\eps\s k) \\
    \M_4&=\gamF(\dop{p'}k\s\eps-\dop{p'}\eps\s k) \,.
  \end{split}\eqlab{gauge-invariants}
\end{equation}
In terms of these operators the difference between
\eqref{cnt-term-2}, generalized to allow
for an admixture of pseudo-scalar and -vector coupling,
and \eqref{cnt-term-1} can be expressed as
\begin{multline}\eqlab{cnt-term-diff}
  \Delta
     = \M_2 \A_2^{pole} (\tF - 1)
     + \M_3 g_{N\Sigma K} \frac{\tf_m(s) + \e\tf_\mk(t)}{m(\chi+1)} \\
     + \M_4 g_{N\Sigma K} \frac{(1-\e)\tf_{m_\Sigma}(u) - \e\tf_\mk(t)}{m(\chi+1)} \,,
\end{multline}
in an obvious notation, and where we have introduced the coupling constant
again. The expression for $\tF$ is given in \eqref{ff-DW} and the $A_2$
amplitude is governed by the convection-current pole contribution,
\begin{equation}\eqlab{A2-pole}
 \A_2^{pole}=\frac{2\e g_\chi}{(s-m^2)(t-\mk^2)}
       -\frac{2(1-\e) g_\chi}{(s-m^2)(u-m_\Sigma^2)} \;,
\end{equation}
with $g_\chi=g_{N\Sigma K} (\chi+(m+m_\Sigma))/2m(\chi+1)$. The model
dependence in the construction of the amplitudes can clearly be seen from
\eqref{cnt-term-diff}. It does not only show in the magnetic
amplitudes ($\A_1$, $\A_3$, and $\A_4$), as is well known, but also in the
convection current, $\A_2$.

As shown by \eqref{cnt-term-diff} the differences between two minimal substitution
procedures can be large. At this point one could take an approach inspired by
that followed in Chiral Perturbation Theory and simply include the most general
structure with parameters which are to be adjusted to the data. For the present
work we opted for a simpler approach where we compare three different
approaches which are often used, namely the Davidson-Workman (DW)~\cite{Dav01}
prescription, the Ohta prescription~\cite{Ohta89}, and the Janssen-Ryckebusch
(JR)~\cite{SJ} prescription as discussed in the following subsections.

In \secref{results} it is shown that the differences between the various
gauge-restoration procedures are important, which should not come as a surprise
since, even at threshold, the energy of the photon is of the same order of
magnitude as the nucleon rest mass. One thus should expect that quark-structure
effects will start to play a role. This short-distance physics is modelled
only very approximately by meson-exchange physics which is the basis of an
effective Lagrangian model.

\subsubsection{The Davidson-Workman prescription}

In our full calculations we have opted to use the DW prescription for the
amplitudes. This prescription implies that for all amplitudes the pole
contributions are taken, modified with form factors appropriate for the
particular diagram, with the exception of the $\A_2$ amplitude which is
modified with an ad-hoc form factor
\begin{equation}\eqlab{ff-DW}
  \begin{split}
    \tF &= \e \tF_{st} + (1-\e) \tF_{su} \\
  \end{split}
\end{equation}
where $\e$ is the charge of the produced kaon, and
\begin{equation}\eqlab{ff-DW-partial}
  \begin{gathered}
    \tF_{st} = F_m(s) + F_\mk(t) - F_m(s) F_\mk(t) \\
    \tF_{su} = F_m(s) + H_{m_\Sigma}(u)- F_m(s) H_{m_\Sigma}(u) \,.
  \end{gathered}
\end{equation}
The full amplitude can now be expressed in the notation
of \eqref{gauge-invariants} as
\begin{gather}
  \begin{split}
    \A_1^{DW}=&\frac{g_\chi F_m(s)}{s-m^2} (1+\kapp) \\
      &+\frac{g_\chi H_{m_\Sigma}(u)}{u-m_\Sigma^2}
         \left( (1-\e)+\frac{m_\Sigma}{m}\kappa_\Sigma \right) \\
      &+\frac{g_{N\Sigma K}} {4m^2(\chi+1)}
         \Bigl(F_m(s) \kapp + H_{m_\Sigma}(u) \kappa_\Sigma \Bigr) \\
    \A_2^{DW}=&\frac{2\e g_\chi \tF}{(s-m^2)(t-\mk^2)}
       -\frac{2(1-\e) g_\chi \tF}{(s-m^2)(u-m_\Sigma^2)}\\
    \A_3^{DW}=&\frac{g_\chi F_m(s) \kapp}{m(s-m^2)} \\
    \A_4^{DW}=&\frac{g_\chi H_{m_\Sigma}(u)\kappa_\Sigma}{m(u-m_\Sigma^2)} \,.
  \end{split}\eqlab{invariant-amplitudes-kaon}
\end{gather}
with $g_\chi=g_{N\Sigma K} (\chi+(m+m_\Sigma))/2m(\chi+1)$,
and $\tF$ is given in~\eqref{ff-DW}.

Originally the DW prescription was presented as an ad-hoc modification
of the convection
current, however following minimal substitution as presented in
\eqref{cnt-term-2} will also lead to the same structure for the amplitude.

\subsubsection{The Ohta prescription}

Following the Ohta prescription~\cite{Ohta89}, which implies minimal
substitution along the lines of \eqref{cnt-term-1} rather than the more
complicated expression of \eqref{cnt-term-2}, the effect
of form factors in the convection current, the $\A_2$ term, is cancelled
completely due to the gauge-restoration procedure. As a result the
$\A_2$ term has no form factors at all $\A_2^{Ohta} =\A_2^{pole}$, as given
in \eqref{A2-pole}. Since the matrix elements of this term are proportional
to the energy, it gives rise to a cross section which increases with energy.
The correction to the convection current of \eqref{invariant-amplitudes-kaon}
can now be formulated as a contact term
\begin{multline}
  \Delta_{Ohta} = (\A_2^{Ohta}- \A_2^{DW}) \M_2 =\A_2^{pole} (1-\tF) \M_2\\
               =2g_\chi \tf_m(s) \left( \e\tf_\mk(t)
                 - (1-\e)\tf_{m_\Sigma}(u)\right) \M_2\,,
\end{multline}
using the notation of~\eqref{ff-twiddle}. Rewriting in terms of $\tf$ shows
clearly that the effect of such a form factor is free from spurious poles.

\subsubsection*{Janssen-Ryckebusch prescription}

In the JR prescription similar form factors are used in the magnetic
current contribution (the contribution to $\A_1$ not proportional to
magnetic moments) as in the convection current. The difference with
the DW amplitude can be written as a contact term,
\begin{multline}
  \Delta_{JR}= (\A_1^{JR}- \A_1^{DW}) \M_1  =g_\chi \Bigl(
          \e F_\mk(t) \tf_m(s) \\
        + (1-\e)(F_m(s)\tf_{m_\Sigma}(u) + F_{m_\Sigma}(u)\tf_m(s))
                \Bigr) \M_1 \,,
\end{multline}
which is free from pole contributions. This prescription was successfully used
in the non-coupled-channel analysis of kaon photoproduction data~\cite{SJ}.

\section{Results \seclab{results}}

The focus of the present work is on strangeness photoproduction. For the
description of the continuum part of the spectrum as well as the width of the
resonances, coupling to other open channels is important. Due to the low
threshold the pion production channel is of particular importance. We will
therefore begin with a short discussion of our pion-nucleon scattering and pion
photoproduction results, to be followed by a discussion of photo-induced kaon
production. We stress that all results are obtained from a single parameter
set.

\begin{figure*}
  \begin{tabular}{cc}
    \includegraphics[angle=90,height=7.9cm]{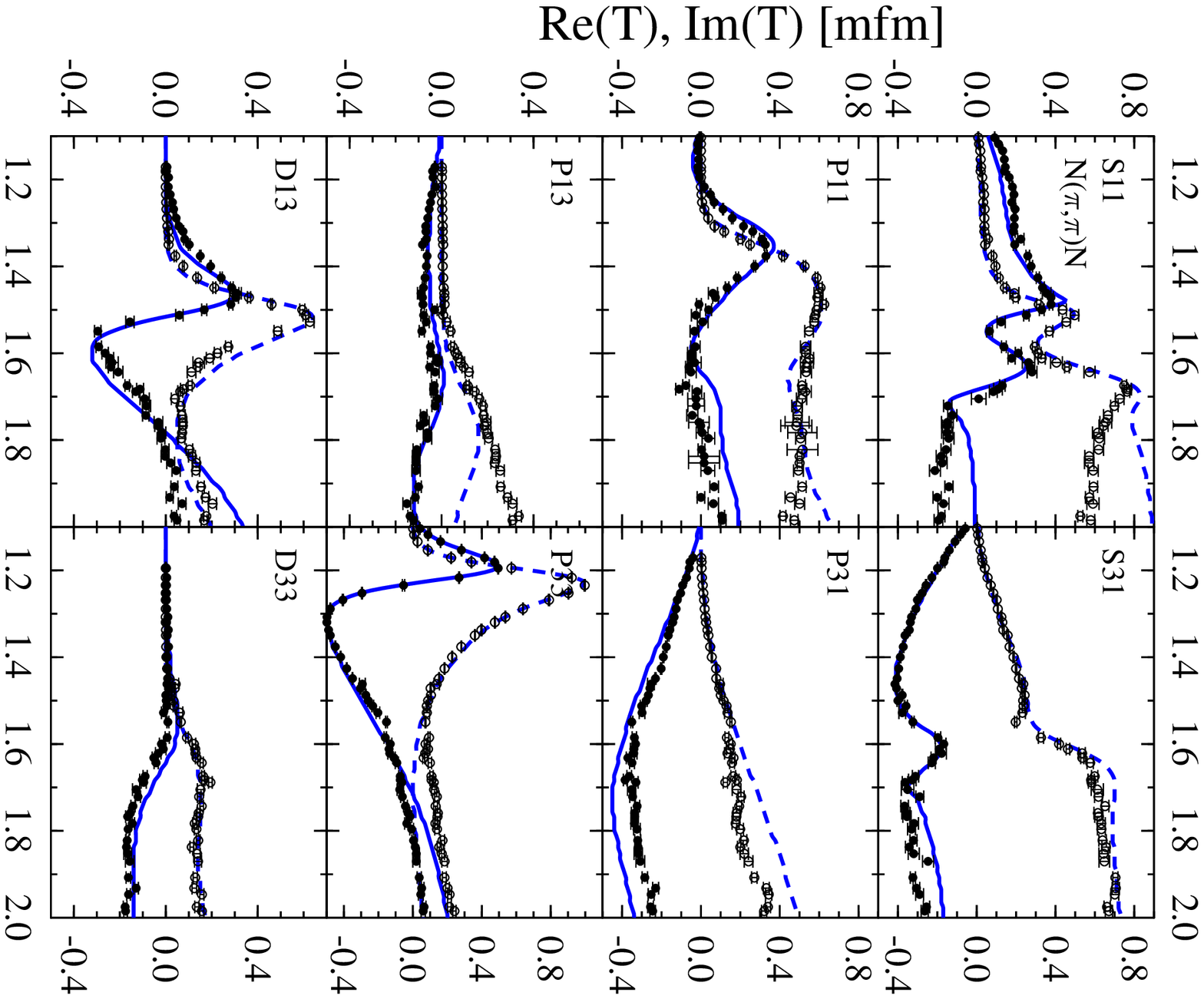} &
    \includegraphics[angle=90,height=7.9cm]{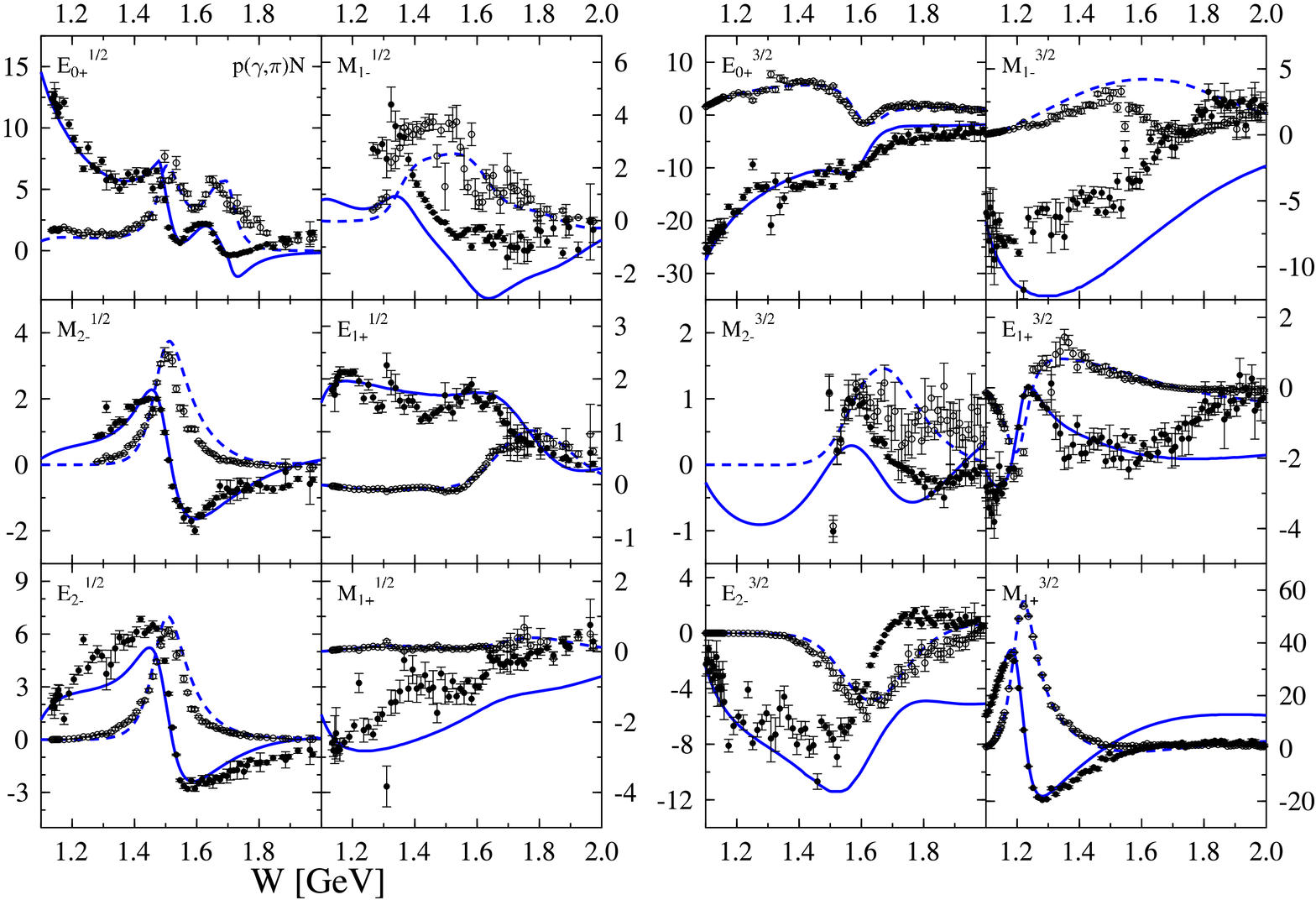}
  \end{tabular}
\caption{Data~\cite{SAID} for pion-nucleon scattering (l.h.s.) and pion
  photoproduction (center and r.h.s.) are compared with the results of our
  calculation. The solid and dashed lines denote the real and imaginary
  parts of the partial-wave amplitudes, respectively.}
  \figlab{pi-pi}
  \figlab{gam-pi}
\end{figure*}

\subsection{$\pi+N \to \pi+N$ and $\gamma+p \to \pi+N$}

The kernel is chosen similar to what had been used in ref.~\cite{Kor98},
with the exception of the contact terms for pion photoproduction, which
are chosen such that the $\A_1^T$ and $\A_2^T$ amplitudes read (only
convection current contributions are shown)
\begin{equation}
  \begin{aligned}
  \A_1^\oh&=\frac{\tF_{su}}{s-m^2} +\frac{\tF_{su} - 2\tF_{ut}/3}{u-m^2} \\
  \A_2^\oh&=\frac{2\tF_{su}}{(s-m^2)(u-m^2)}
       -\frac{4\tF_{ut}/3}{(u-m^2)(t-m_\pi^2)} \\
  \A_1^\th&=\frac{\tF_{ut}}{u-m^2} \\
  \A_2^\th&=\frac{2\tF_{ut}}{(u-m^2)(t-m_\pi^2)} \,,
  \end{aligned}
\end{equation}
where $G_{su}$ and $G_{ut}$ are defined similar to \eqref{ff-DW-partial}.
This improves the amplitudes at higher energies, while the low-energy
behavior is not affected.
The results of the calculations are compared with the partial-wave
data obtained from the analysis of the Virginia Polytechnic Institute (VPI)
group~\cite{SAID} in \figref{pi-pi}. The experimental data
are reproduced with reasonable accuracy up to the energies of 1.7-1.8 GeV.
In the $\pi-N$ sector we attribute the discrepancies at high
energies primarily to the inelastic channels not explicitly included
in the model. An exception is the \res{P}{11} partial wave, which is
traditionally problematic. In our calculation its large width is generated
partly due to a large pion-nucleon coupling and partly because of a
large decay width to the two-pion production channel.

In the pion photoproduction amplitudes the largest discrepancies are
seen in the $\M_{1-}$ and $\E_{2-}$ partial waves for isospin $3/2$. Our
investigations show that the $\E_{2-}$ partial wave is almost exclusively
sensitive to the $\A_2$ contribution, and can be corrected using
form factors with cut-off values well below 1 GeV. While this will
improve the $\E_{2-}$ partial wave, it also influences strongly
all other multipoles and we have chosen not to adapt this procedure for
our final calculations. In the case of $\M_{1-}$ partial wave, in addition
to the convection current, a number of other sources contribute strongly,
such as magnetic terms from the Born contributions and the
$\omega$ t-channel. All these contributions are however fixed by their
components in other partial waves.

To be able to estimate the effects of the missing inelastic channels on the
kaon sector we introduced in the model the $\rho-N$ final state where the
coupling is through Born terms only. As expected, its inclusion generates
sizable in-elasticities for certain multipoles in the $(N+\pi \to N+\pi)$
sector.

We note, that by fitting the pion scattering and pion photoproduction
amplitudes, masses, pion- and photon-coupling constants for the most of
the resonances are fixed which strongly limits the number of free
parameters for the kaon-production channels.

\begin{figure*}
  \begin{tabular}{c@{}c}
    \includegraphics[angle=90,height=7cm]{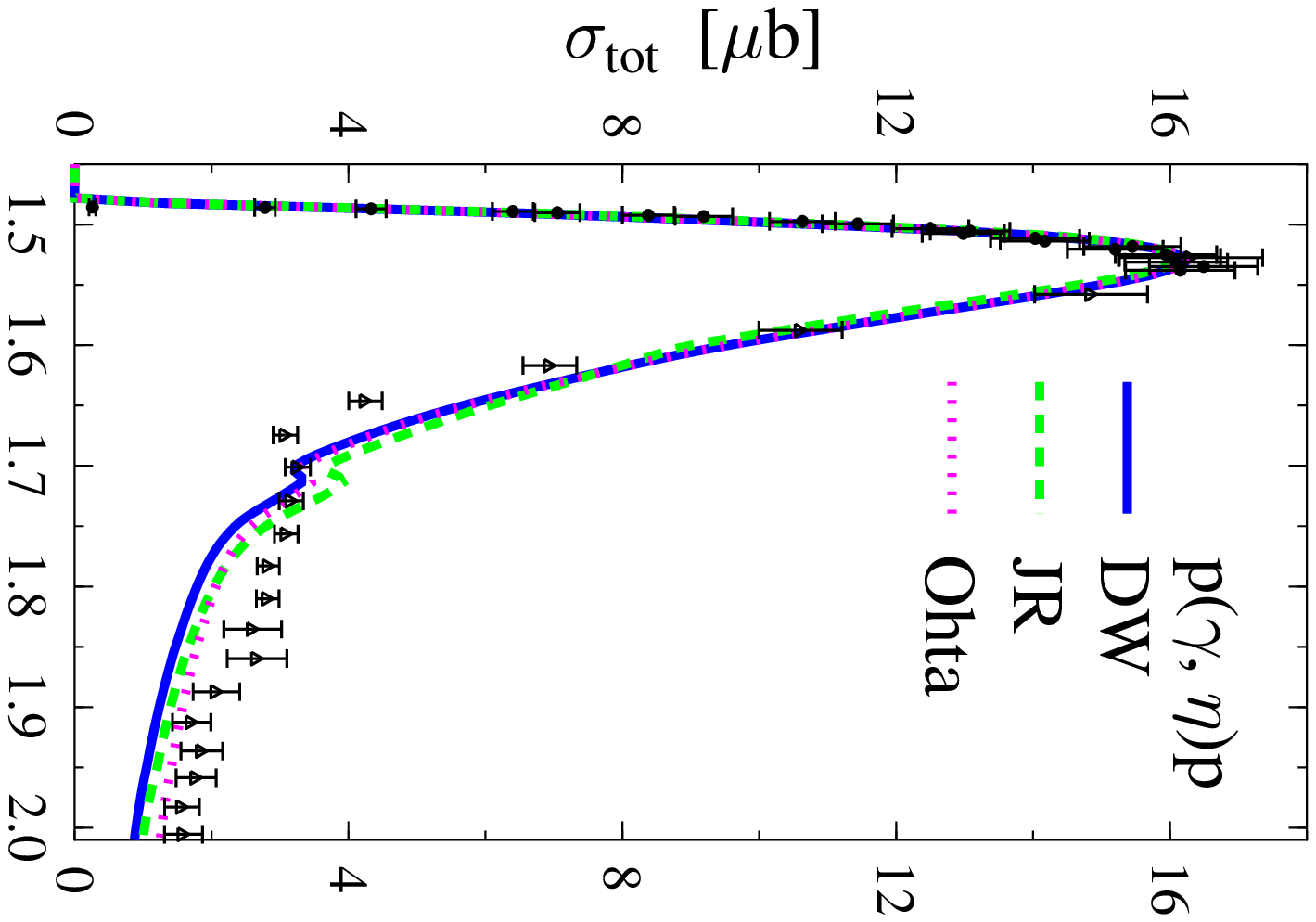} &
    \includegraphics[angle=90,height=7cm]{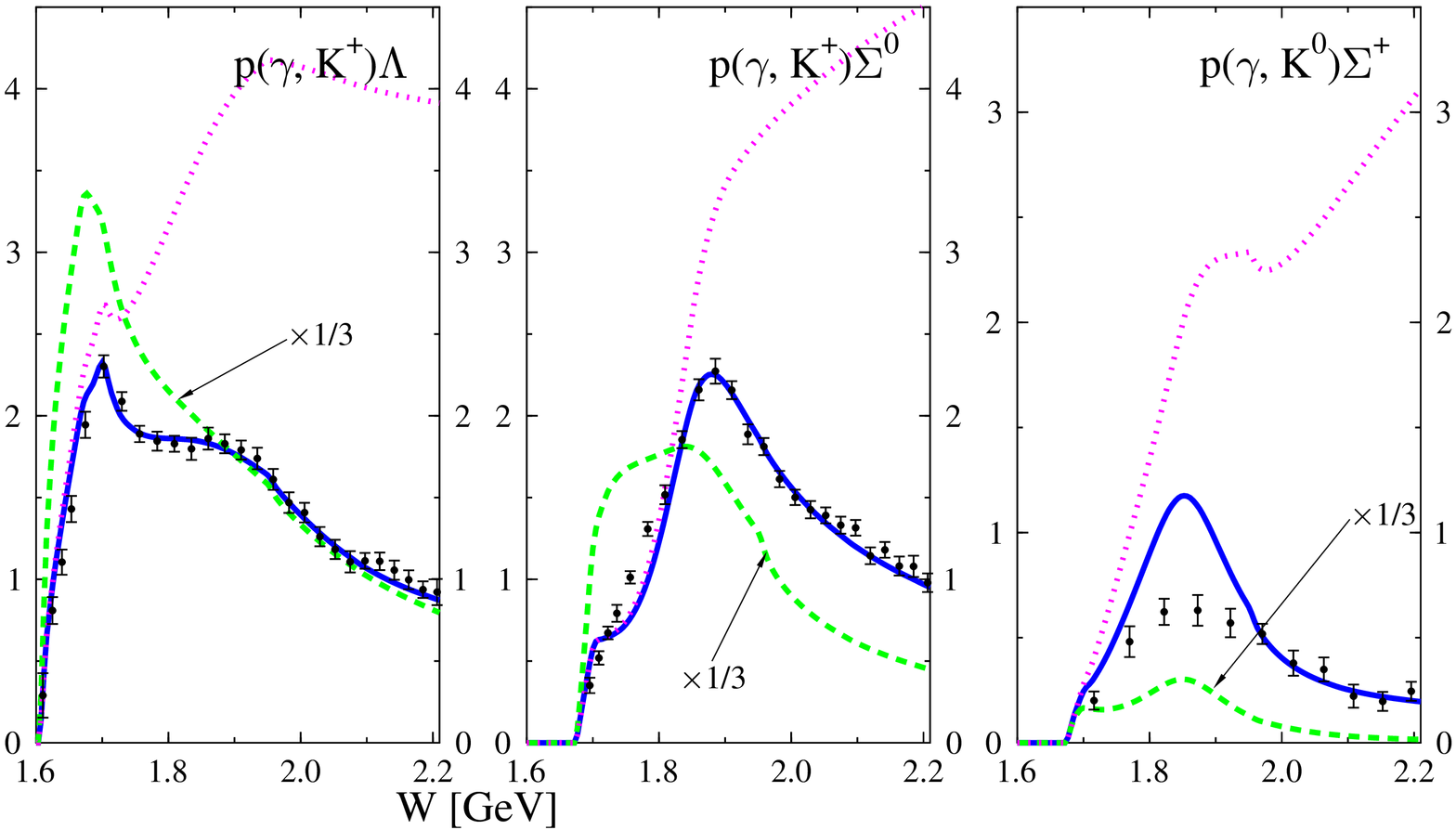}
  \end{tabular}
  \caption{[Color online] Total cross-sections for $\eta$ and kaon photoproduction.
    Solid line corresponds to the current fit, based on the DW prescription
    for gauge invariance restoration. Dashed line -- same parameters but
    using a Janssen - Ryckebusch (JR) prescription. Dotted line --
    same using the Ohta prescription. Note that the JR calculation in
    kaon sector is scaled down by a factor of 3.
    The data has been taken from refs.~\cite{SAPHIR1,SAPHIR2,SAID,eta2}.}
  \figlab{eta-tot}
  \figlab{kl-tot}
  \figlab{ks-tot}
\end{figure*}

\subsection{$\gamma+p \to \eta+p$}

The data for the ($\gamma + p \to \eta + p$) reaction is taken from the on-line
database of the VPI group~\cite{SAID} and from~\cite{eta2}. Our calculation
reproduces the cross-section for this reaction channel rather well, see
\figref{eta-tot}. As is well known the \rres{S}{11}{1525} and
\rres{S}{11}{1690} resonances give the major contribution to the cross section.
In the following section it is shown that  re-scattering via the $(\gamma + p
\to \eta + p)$ channel influences the kaon production channels. It is worth to note,
that the effects due to the different choices for the gauge restoration procedure
in the $\eta$-channel are negligible, primarily due to the dominance of the resonance
contributions in this channel and the relative weakness of the Born contributions.

\begin{figure*}
  \includegraphics[angle=90,width=1.0\textwidth]{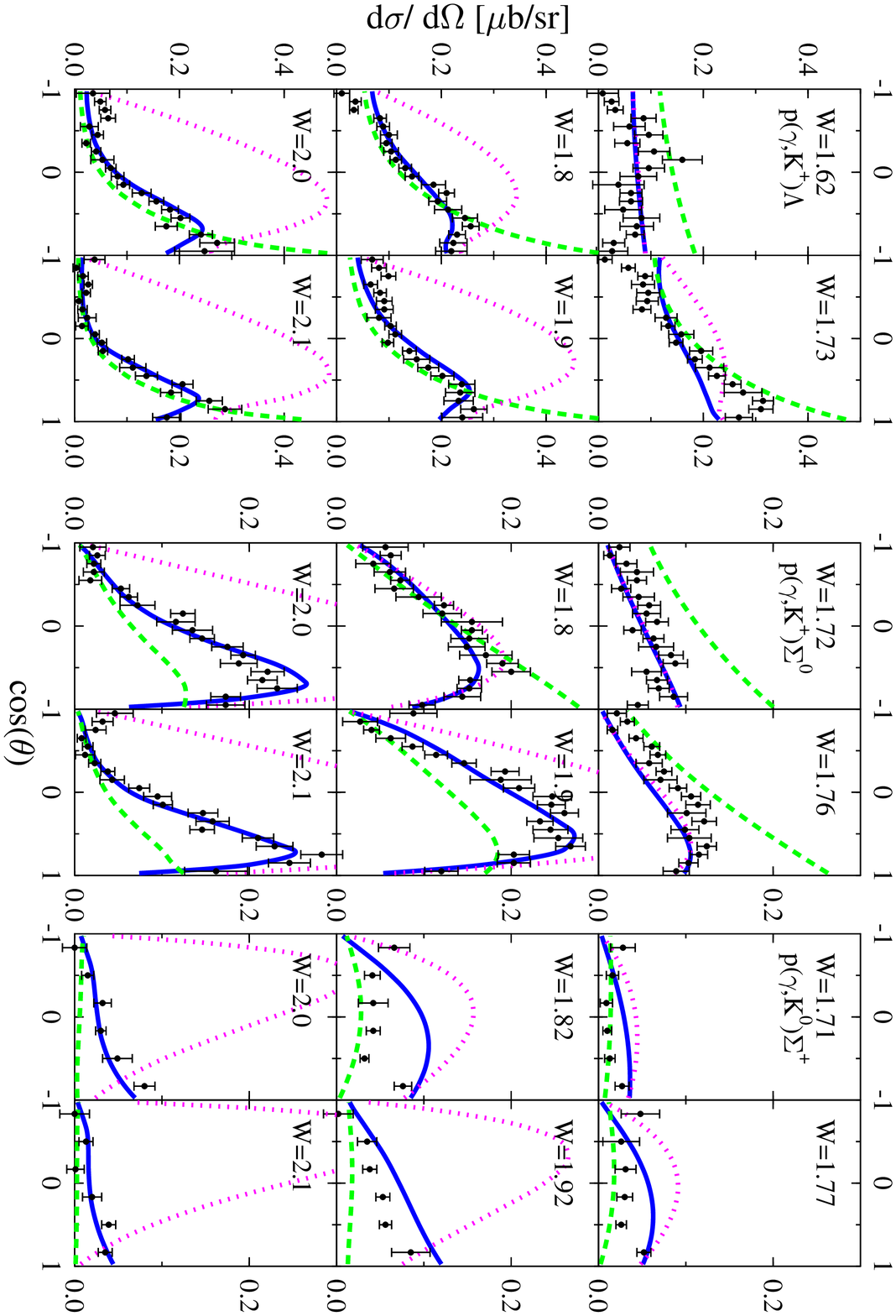}
  \caption{[Color online] Differential cross sections for kaon photoproduction.
    Line coding is the same as in \figref{kl-tot}. The experimental data
    are from the SAPHIR collaboration~\cite{SAPHIR1,SAPHIR2}.}
  \figlab{kl-dif}
  \figlab{ks-dif}
\end{figure*}

\begin{figure*}
  \includegraphics[angle=90,height=7cm]{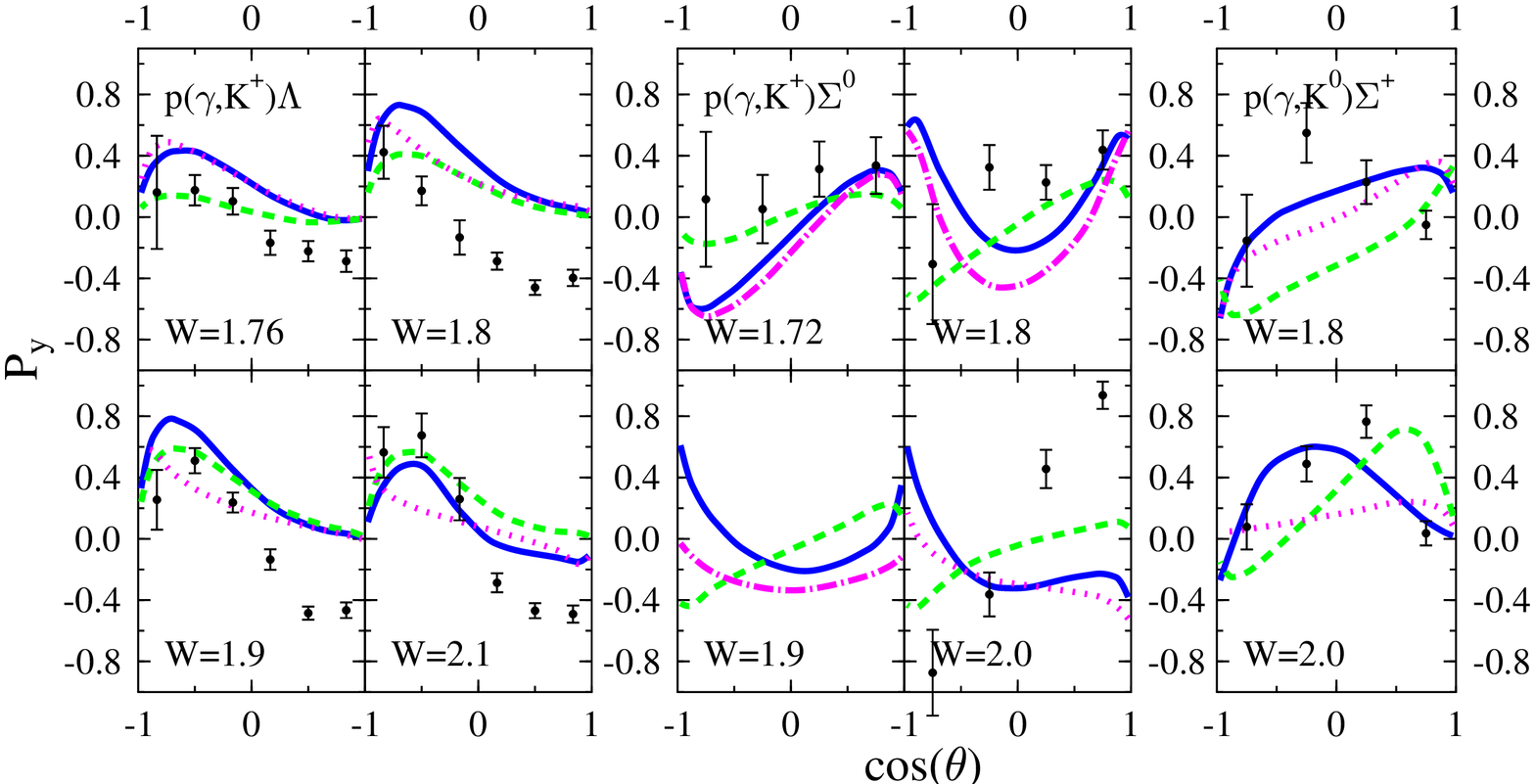}
  \caption{[Color online] Same as \figref{kl-dif} for the final-state polarization in
    kaon photoproduction.}
  \figlab{kl-pol}
  \figlab{ks-pol}
\end{figure*}

\subsection{$\gamma + p \to K^+ +\Lambda$}

Our results for photo-induced kaon production are compared to the data from the
SAPHIR collaboration~\cite{SAPHIR1} for angle-integrated cross sections in
\figref{kl-tot}, for angular distributions in
\figref{kl-dif}, and for final-state polarization in
\figref{kl-pol}. The overall agreement with the data is very good, including
the asymptotic behavior of the cross-section at high energies. The peaking at
90$^\circ$ seen in the differential cross-section at threshold is not
reproduced by our calculation. It should however be noticed that in the CLAS
data~\cite{CLAS1}, see~\secref{results-CLAS}, there is ample evidence for a
peaking at 90$^\circ$ at the lowest energies.

The results for the $(\gamma + p \to K+\Lambda)$ cross-sections strongly depend
on the choice for the gauge restoration prescription used. In the case of the
Ohta prescription the $\A_2$ amplitude is unquenched in the high-energy limit,
resulting in an unbounded growth of the cross-section. The use of the JR
prescription, as compared to the DW, results in a much larger value for the Born
contributions. The choice for the gauge restoration procedure is also important
for the angular distributions; the use of the JR prescription results in a much
more forward-peaked differential cross-section as compared to the prediction
following from the DW prescription.

While for the $K + \Lambda$ photoproduction channel it is possible to fit the
experimental data using both prescriptions, we prefer the DW prescription for
the $K + \Sigma$ photoproduction channel. The couplings extracted using the
DW prescription are rather close to the $SU(3)$-model predictions, see
\tblref{summary-BBP}. For the JR prescription we would have to suppress the
$g_{N\Lambda K}$ coupling by a factor of $\sqrt3$ to compensate the
enhancement of the background contributions, which will contradict the
predictions of the $SU(3)$ model.

\begin{figure}
  \includegraphics[width=1.0\linewidth]{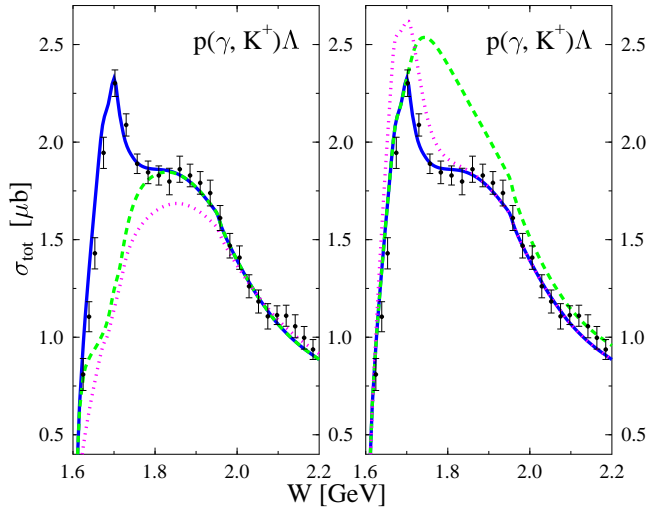}
  \caption{[Color online] Illustration of the re-scattering effects where the solid
    curve (corresponding to our full calculation) is the same for the
    left and right panels and serves as a reference. See the text for
    explanation for the other curves. Note, that the zero point is
    suppressed in these plots.}
  \figlab{kl-resc}
\end{figure}

The coupled-channel effects are large in the $K-\Lambda$ channel, it
not only gives an overall depletion of the cross section at higher
energies, but it is responsible for certain structures seen in the
spectrum. In particular we find that the narrow structure at 1.7 GeV
is generated through re-scattering effects. To illustrate this the
results of the following benchmark calculations are compared in
\figref{kl-resc}.
\begin{itemize}
\item The dashed line in the l.h.s.\ panel corresponds to a calculation
  in which the photon-nucleon coupling constant for the \rres{S}{11}{1690}
  resonance, $g_{N\gamma}$, is decreased by a factor 10 and the kaon coupling,
  $g_{K\Lambda}$, increased by the same factor such, that its leading order
  contribution to the $(\gamma+p \to K+\Lambda)$ reaction remains unchanged.
  While
  the full calculation shows the peak at 1.7 GeV, it is  missing in the
  dashed calculation which shows that it is not due to a direct contribution
  from the resonance, but that re-scattering effects are essential in
  its formation.
\item A detailed investigation at the level of partial wave amplitudes also shows
  that the peak is due to indirect (coupled-channels) contributions.
  The direct contribution from the $(\gamma+p \to \res{S}{11} \to K+\Lambda)$
  diagram is almost completely cancelled by the corresponding one
  from the $(\gamma+p \to \pi + N \to \res{S}{11} \to K+\Lambda)$ channel
  coupling. To illustrate the complicated cancellation we show as the dotted
  line the results of a calculation where $g_{NK^*\Lambda}$ coupling constant
  is set to zero. Combined with a results of the dashed curve this shows that
  the dominant contribution to the peak is through the coupled channels
  contribution $(\gamma+p \to \res{S}{11}\to \pi + N \to  K+\Lambda)$,
  \m{where the $K^*$ t-channel contributes to the last step.
  The direct $K^*$ t-channel contribution affects the partial wave
  amplitudes for $(\gamma+p \to  K+\Lambda)$ channel, but the net
  result on the cross-section is small.}
\item The dashed line in the r.h.s.\ panel of \figref{kl-resc} represents
  the results of a calculation in which the $\rho$-meson final state has been
  excluded from the model space. The opening of the $\rho$-meson channel takes
  flux away from the   $K\Lambda$-channel thus depleting the cross section near
  the threshold.
\item The dotted line corresponds to a calculation in which the signs of
  the $\eta$ coupling-constants for both \res{S}{11} resonances have been
  changed. In leading order this only changes the interference pattern for
  the $N - \eta$ final state and has no effect on   any other channel. As
  can be seen the effects of channel-coupling are appreciable, even for a
  rather subtle change in coupling constants in other channels.
\end{itemize}

\begin{figure}
  \includegraphics[angle=90,width=1.0\linewidth]{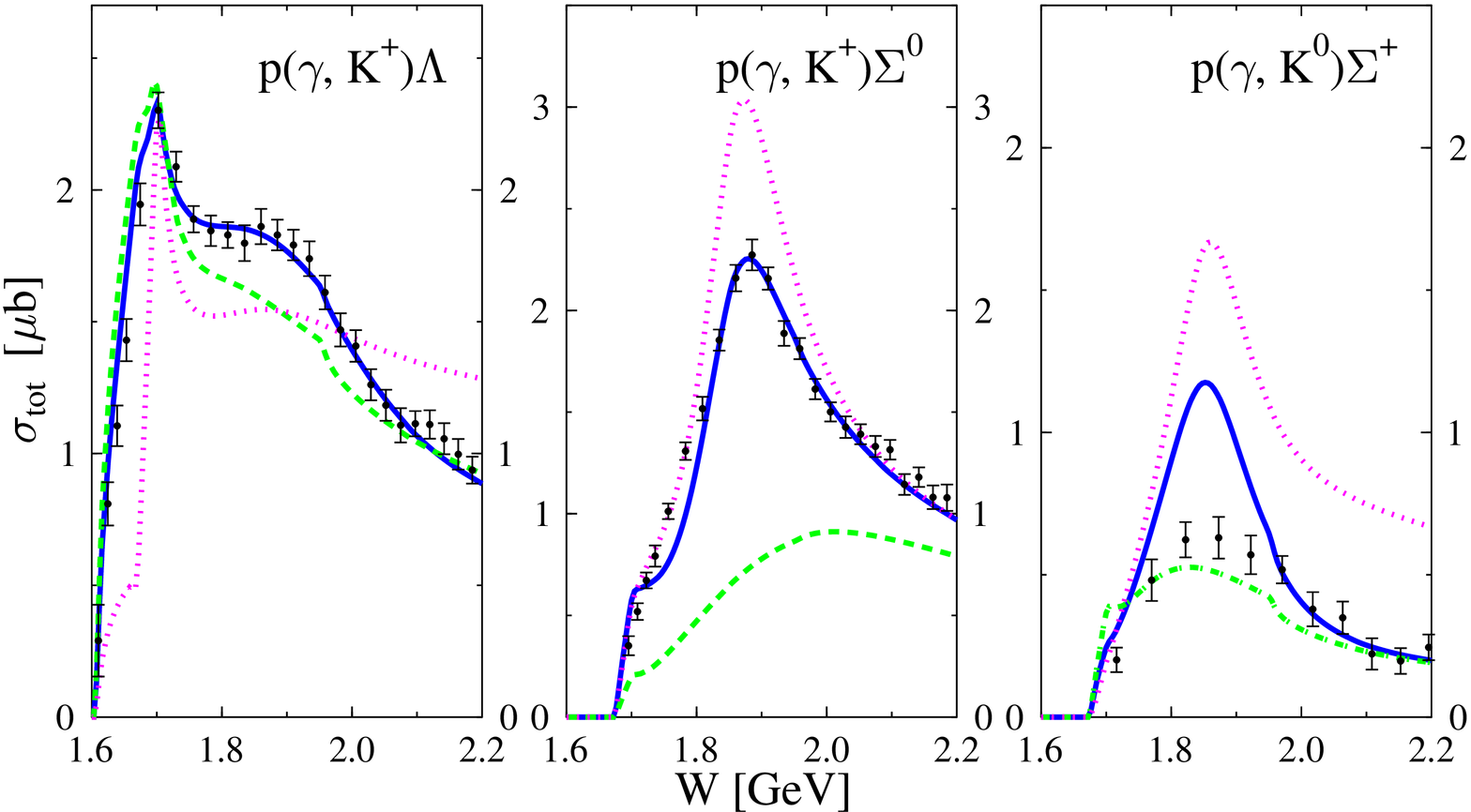} \\
  \includegraphics[angle=90,width=1.0\linewidth]{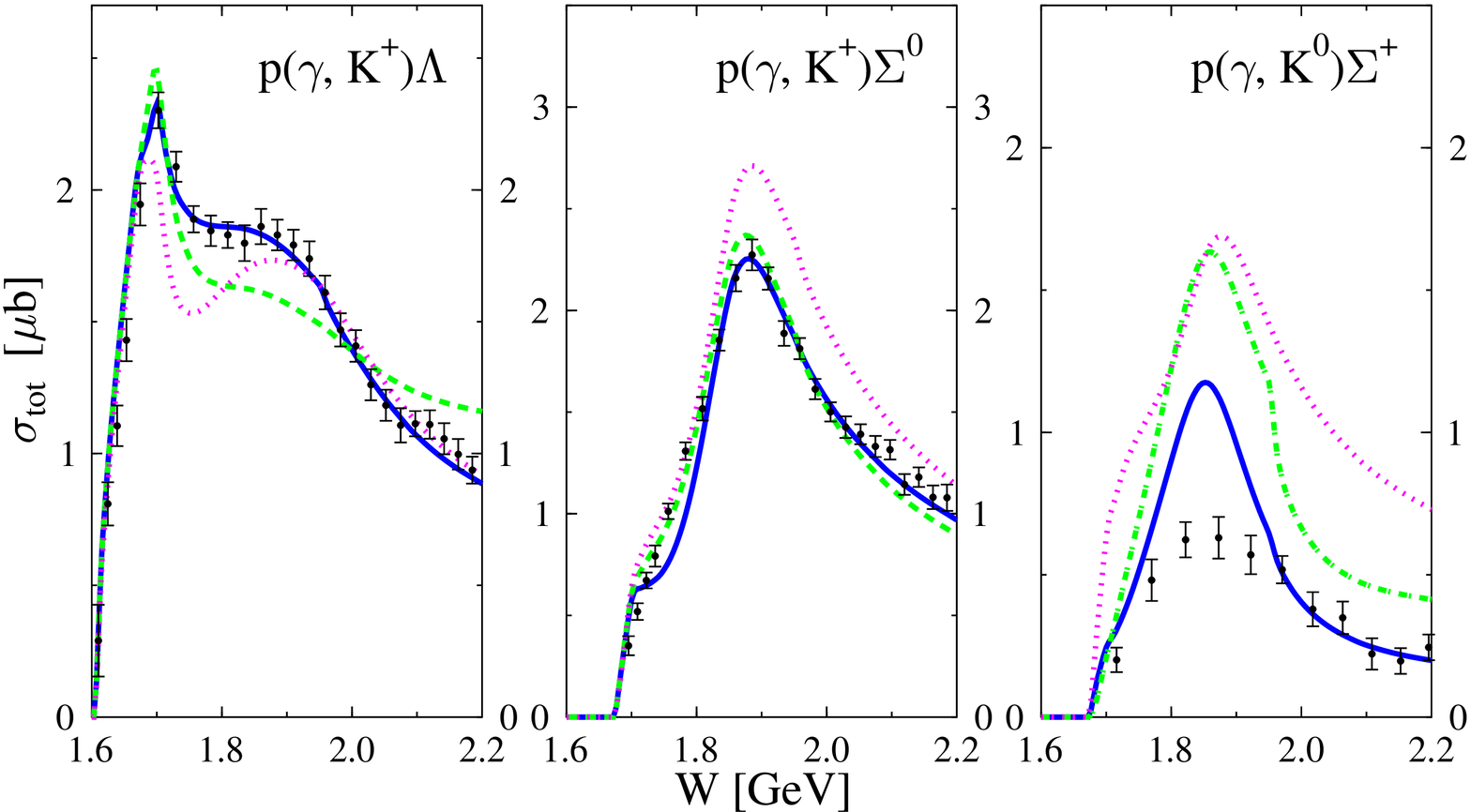}
  \caption{[Color online] The importance of the different contributions for
strangeness production is shown. The solid line corresponds to the full
calculation. The dashed line in the upper plane corresponds to the calculation
where the resonance contributions in the kaon photoproduction channels have
been excluded. The dotted line corresponds to the calculation where the
channel-coupling effects were switched off. The dashed line in the lower plane
corresponds to the calculation, where the $\rho$ t-channel contributions
between the different kaon final states were switched off. The dotted line
shows the results of a calculation using dipole form-factors \eqref{ff-dipole}
in the u-channel instead of those of \eqref{ff-u-channel}.}
  \figlab{k-resonances}
\end{figure}

Our calculations show, as illustrated in \figref{k-resonances}, that a major
part of the cross-section is generated via non-resonant and re-scattering
contributions. Even some prominent structures in the spectrum can be understood
as arising through a coupled channels effect. Resonances give only a relatively
minor direct contribution to the cross section in the $(\gamma + p \to
K+\Lambda)$ channel. The sharp peak in the tree-level calculation corresponds
to the second \res{S}{11} resonance and is so prominent due to its small width.
In a coupled-channels calculation the width is increased due to the coupling
to the $\eta$ channel.
\m{The figure shows that for $\Lambda$ production the structure of the
u-channel form factor in the strange-sector is not very important. The reason
is that the u-channel contribution with an intermediate $\Sigma$ or
$\Lambda$-baryon gives only a minor contribution to the cross section. The use
of the form factor of \eqref{ff-u-channel} gives rise to a stronger suppression
of the u-channel contributions than the usual dipole form~\eqref{ff-dipole},
see~\figref{k-resonances}. }

\begin{figure}
  \includegraphics[angle=90,width=1.0\linewidth]{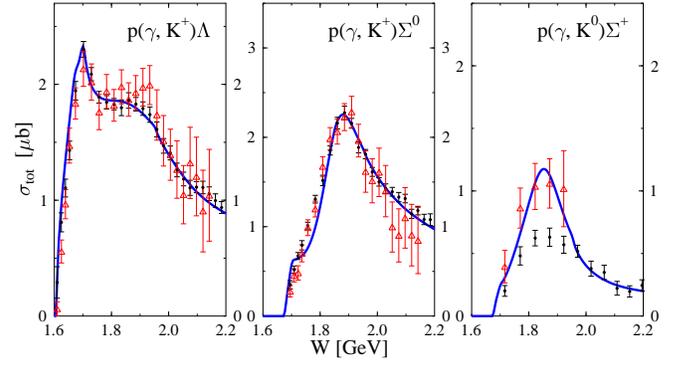}
  \caption{[Color online] Comparison of the old~\cite{SAPHIR1-old,SAPHIR2-old}
and new~\cite{SAPHIR1,SAPHIR2} data for angle integrated cross
sections from the SAPHIR Collaboration.}
  \figlab{saphir-old-new-1}
\end{figure}

\begin{figure*}
  \includegraphics[angle=90,width=.8\linewidth]{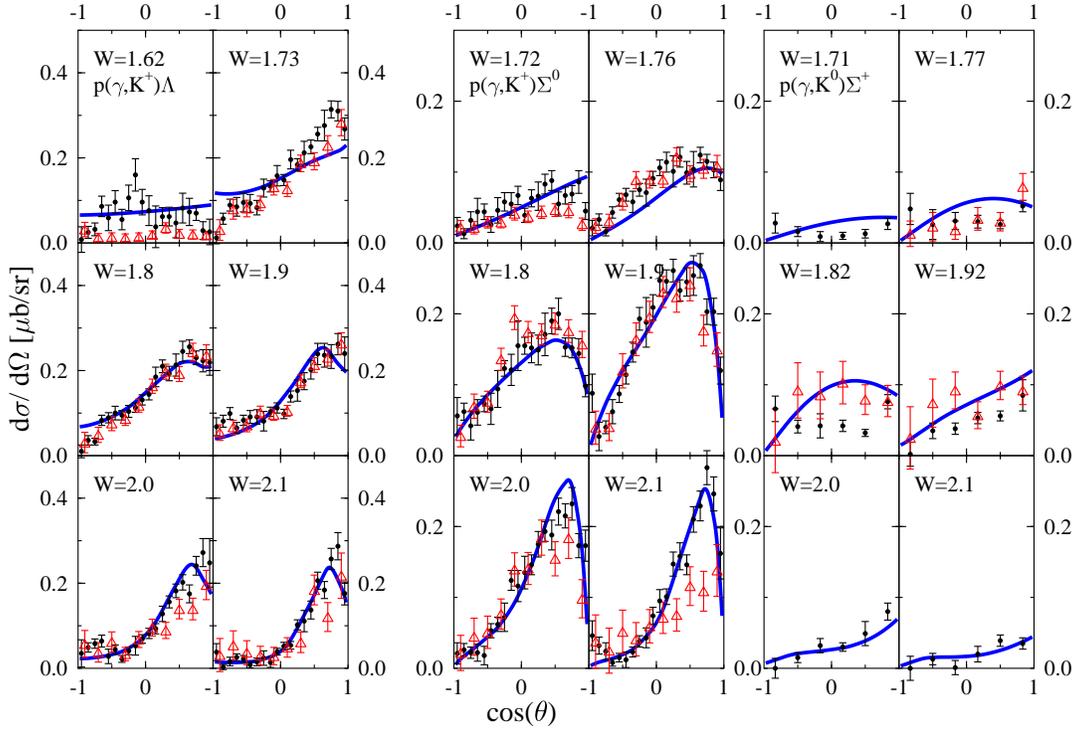}
  \caption{[Color online] Comparison of the old~\cite{SAPHIR1-old,SAPHIR2-old}
and new~\cite{SAPHIR1,SAPHIR2} data for differential cross sections from the
SAPHIR Collaboration (triangles).}
  \figlab{saphir-old-new-2}
\end{figure*}

\begin{figure}
  \includegraphics[width=1.0\linewidth]{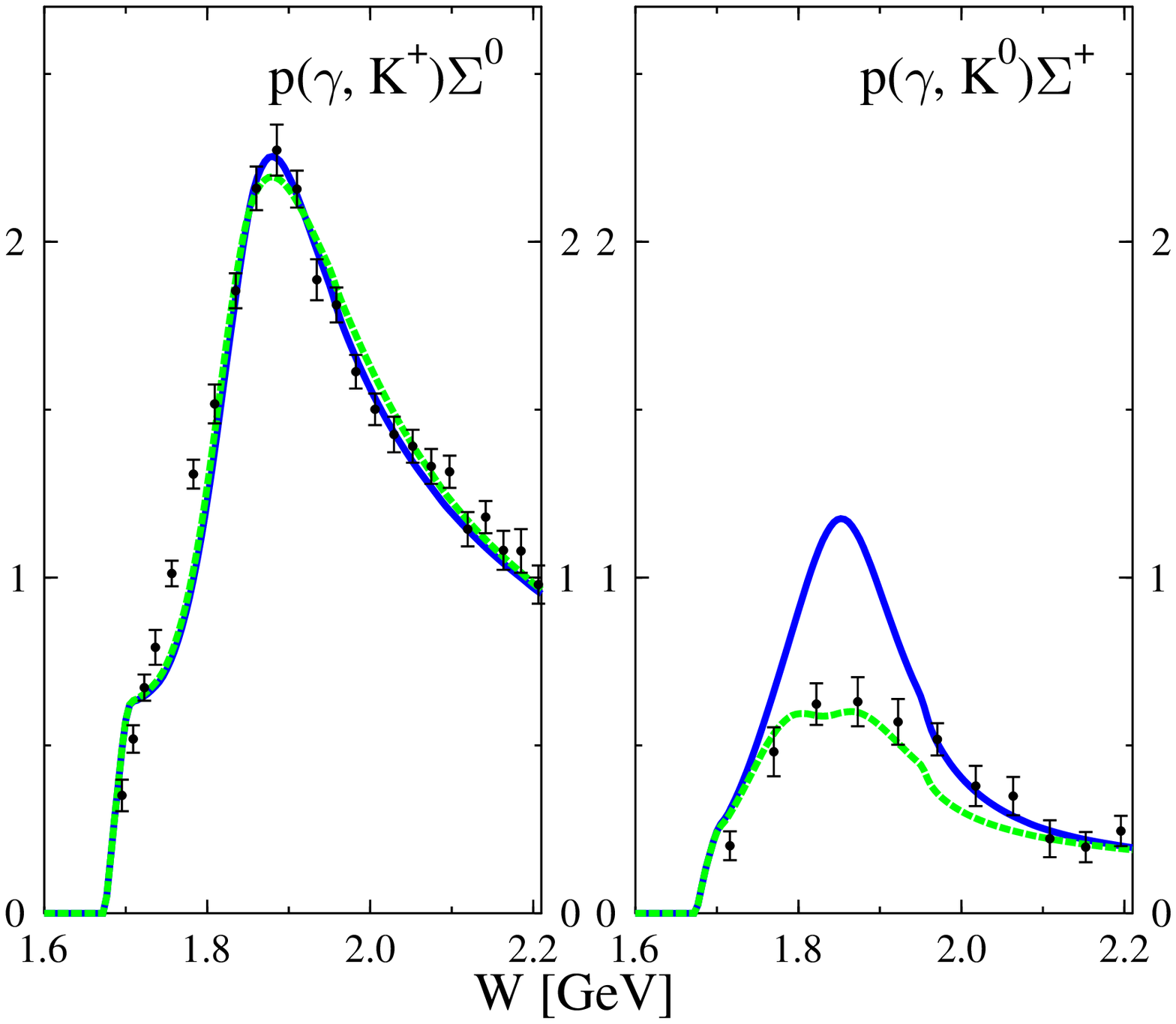}
  \caption{[Color online] Illustration of the effects of the introduction of
    a second \res{P}{13} resonance.}
  \figlab{p13b}
\end{figure}

\subsection{$\gamma+p \to K^++\Sigma^0$ and
  $\gamma+p \to K^0+\Sigma^+$ \seclab{results-KS}}

The data from two different analyses by the SAPHIR group are compared in
\figref{saphir-old-new-1} and \figref{saphir-old-new-2}. While for $\Lambda$-
and $\Sigma^0$-production the two analyses basically agree, they show big
differences for $\Sigma^+$-production. The low value for the
$\Sigma^+$-production cross section poses a problem for the interpretation of
the data since the isospin Clebsch-Gordon coefficients are in general
larger for $\Sigma^+$ than for $\Sigma^0$ production.  In the $K^+\Sigma^0$
channel the new data shows a stronger forward peaking of the cross section at
high energies than the old.

The difference between the DW and JR prescriptions in the $\Sigma$ channel is
even more pronounced than in the $\Lambda$ channel, see \figref{ks-tot} and
\figref{ks-dif}. The use of the JR prescription results in a larger value for
the non-resonant contribution in the $K^+\Sigma^0$ channel while that for the
$K^0\Sigma^+$ channel is depleted, more in agreement with the new data.
However it also predicts a strongly forward-peaked angular distribution in the
$\Sigma^0$ channel and little forward peaking in the $\Sigma^+$ channel, in
disagreement with the data.

In the case of the DW prescription the ratio $\Sigma^0/\Sigma^+$ is much smaller
than for JR. To correct for this we introduced modified u-channel form-factors
and included $\rho$-meson t-channel contributions. The effects of these
contributions is illustrated in the lower plane of \figref{k-resonances}. For
the dashed line the $\rho$-meson t-channel contribution to the different matrix
elements for the kaon states is switched off, which enhances the cross section
in $K^0\Sigma^+$ channel and depletes that for the $K^+\Lambda$ channel. The
basis for this is a relatively subtle interference between the channels. In
obtaining the dotted line the usual dipole form for the u-channel form-factors
has been taken. This results in a much larger cross-section especially at
higher energies \m{which is in disagreement with the data. The main effect of
the use of the form factor \eqref{ff-u-channel} is to suppress the contribution
of diagram with a $\Sigma$-baryon in the u-channel and thus suppressing the
cross-section at backward angles for $\Sigma^+$ production.}
In the full calculation
the cross section at higher energies agrees with the data and shows a gradual
decrease at energies beyond those plotted. In $\Sigma^0$-production an enhanced
cross section at 90$^\circ$ at around 1.9 GeV is observed, which is a strong
sign for a $P$ or $D$ resonance, where in the present calculation it is
explained via a \rres{P}{33}{1855} contribution. However, since such a
resonance also contributes strongly to the $K^0\Sigma^+$ channel, due to
isospin symmetry, it needs to be compensated there through the introduction of
an additional \res{P}{13} resonance at a similar energy. With the correct
couplings the two resonances will interfere constructively in the
$K^+\Sigma^0$ channel and destructively in the $K^0\Sigma^+$ channel.
\m{To illustrate this point the results of a calculation is shown in the
\figref{p13b}, in which a \res{P}{13} resonance is included  at
$m_{P_{13}}=1830$~MeV with a coupling strength of $g_{{P_{13}}K\Sigma}=0.05$
and width of $W_{P_{33}}=0.10$~GeV, at the same time increasing the width of
${P_{33}}$ to $W_{P_{33}}=0.25$ (using the same photon couplings as shown in
\tblref{resonances} for both of them). Shown are
total cross-sections for $K^0\Sigma^+$ and $K^+\Sigma^0$ channels.
The angular distributions for the $K^+\Sigma^0$ channel are also improved, while
those for $K^0\Sigma^+$ are not affected.
The rest of observables (including final-state polarization) are not modified
or modified only slightly.}

Compared to the $K^+\Lambda$ channel the resonances play a larger role in
the $\Sigma$ channels, especially in the $K^+\Sigma^0$. At the same time we
have found the inclusion of the coupled-channels effects to play an important
role, as it suppresses the cross-section at the higher energies.

\begin{figure*}
  \includegraphics[angle=90,width=.8\linewidth]{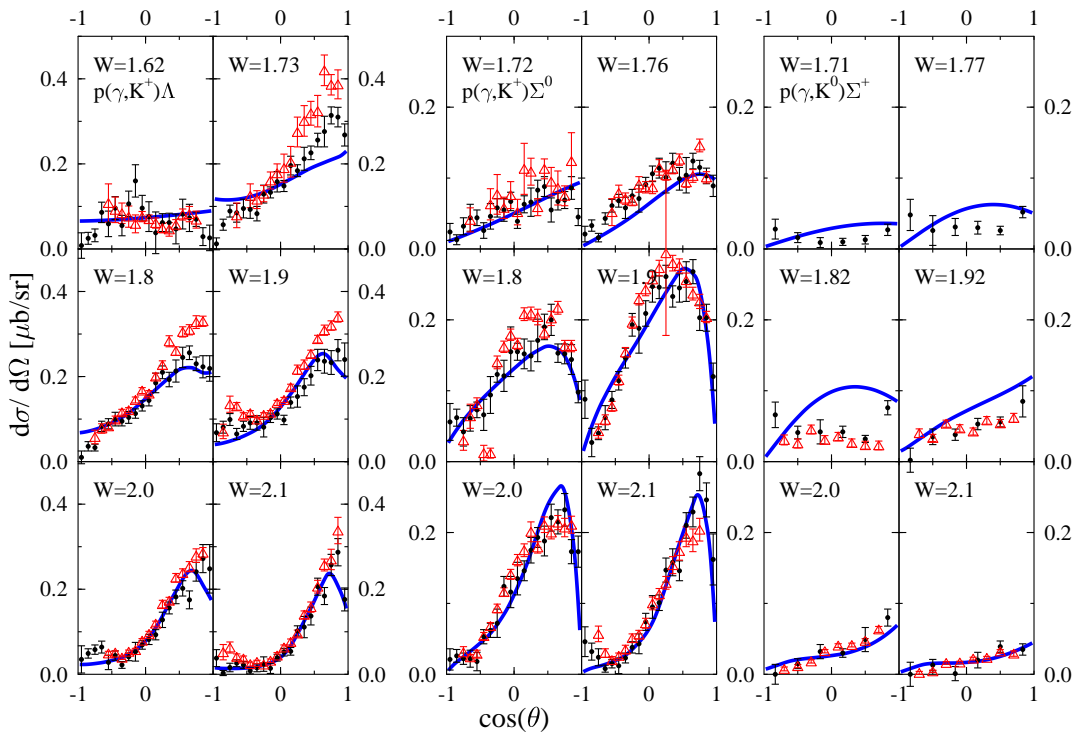}
  \caption{[Color online] Comparison of our calculation with the new data from
    the SAPHIR Collaboration~\cite{SAPHIR1,SAPHIR2} (circles) and the data
    from the CLAS collaboration~\cite{CLAS1,CLAS2} (triangles).}
  \figlab{saphir-clas}
\end{figure*}

\subsection{Comparison with the CLAS data \seclab{results-CLAS}}

In \figref{saphir-clas} the new SAPHIR data~\cite{SAPHIR1,SAPHIR2} is compared
to the data from the CLAS collaboration~\cite{CLAS1,CLAS2}. From this figure we
can see that \m{the two data sets agree in overall magnitude but do show some
important differences.}

In the $K^+\Lambda$ channel the most important difference between the two data
sets is the cross-section at the very forward angles. For the SAPHIR data, the
cross-section clearly drops off, while this is not
seen in the CLAS data. As it was noted before, the trend at forward angles is
very sensitive to the gauge restoration scheme chosen.
\m{
An alternative way to account for this difference will be the inclusion of the
$K_1$ t-channel contribution~\cite{Ire04}, which is not needed for the SAPHIR
data. } Another difference lies at backward angles where the CLAS data at 1.9
GeV and 2.1 GeV show a much more pronounced peak in the angular distribution.
Our calculation does not reproduce this peak, but it could possibly be the
indication for an additional $P$ or $D$ resonance.

In the $K^+\Sigma^0$ channel the difference between the two data sets is that
in the CLAS data the bump in the cross section at 90$^\circ$ is slightly more
pronounced than in the CLAS data.

\section{Conclusions}

In this work we showed that for a calculation of meson production at higher
energies, in particular kaon production, it is essential to perform a full
coupled channels calculation. The effects of channel coupling are not just a
smooth change of the energy dependence of the cross section, but can also
give rise to structures in the cross section which might otherwise be
misinterpreted as resonances.

An additional advantage of a full coupled channels calculation is that it
allows for a simultaneous calculation of observables for a large multitude of
reactions with considerably fewer parameters than would be necessary if each
reaction channel would be fitted separately. As shown the coupled-channels
calculation is manageable if the K-matrix formalism is used.

For photoproduction reactions it is necessary to perform a gauge-invariant
calculation. The particularities of the gauge restoration procedure are however
model dependent. For low photon energies this model dependence does not reflect
strongly on observables. At higher photon energies ($\omega$), corresponding to
the threshold of kaon production and beyond, the model dependencies in gauge-restoration
procedures give rise to strongly different Born contribution to the amplitude
and result in large differences in extracted coupling parameters or predictions
for cross sections that are at variance with the data. The reason for this is
that the differences between the various gauge-restoration procedures are of
higher order in $\omega$ which gives rise to large terms when
$\omega/m\approx 1$. The observables are thus sensitive to short-range effects
and one is entering the energy regime where quark effects will start to be
important.

The calculations presented in this work show that a good fit to the data can be
obtained using the DW prescription for the gauge restoration terms. The model
parameters were fitted to the data and are largely consistent with SU(3). Some
notable differences from SU(3) and some discrepancies in reproducing the data
could be attributed to the need for a second $\res{P}{13}$ resonance at an
energy of around 1.9 GeV.

\begin{acknowledgments}
This work was performed as part of the research program of the Stichting voor
Fundamenteel Onderzoek der Materie (FOM) with financial support from the
Nederlandse Organisatie voor Wetenschappelijk Onderzoek (NWO). We thank
A.Yu.~Korchin, R.~Timmermans and T.~Corthals for discussions and a
critical reading of the manuscript.

\end{acknowledgments}

\appendix

\section{Lagrangian}
\label{app:lagrangian}

In this section we present the effective-field Lagrangian used in this
work. A comparison is presented with $SU(3)$ predictions of the model
parameters and their values as extracted from our fit. A summary of the $SU(3)$
notation used is presented in Appendix~\ref{app:su3}.

The $SU(3)$-model predictions for the coupling constants of baryons have
a long history dating back to the papers by Rijken and
de~Swart~\cite{Swa79}, where they found $F/D \simeq 0.69$.
In ref.~\cite{Swa89} a value of $F/D \simeq 0.55$ was found.
In ref.~\cite{Bor99} a recent calculation
is presented on the extraction of $SU(3)$-model parameters from semi-leptonic
decays $B \to B'+e^- +\bar\nu_e$. The extracted values range from
$F/D \simeq 0.84$ down to $F/D \simeq 0.59$. A calculation based on QCD sum
rules~\cite{Kim00} arrived at a rather small value $F/D \simeq 0.2$.
The values of the couplings constants obtained with $F/D=0.55$ in the extreme
quark model ($S=3F-D$) are presented in a~\tblref{summary-BBP} (BBP, upper plane)
and agree very well with the values we obtained from our fit with the
only notable exception of $g_{N\Sigma K}$. The $SU(3)$ prediction for this
coupling constant is very sensitive to the exact value of $F/D$ and our
extracted values can be seen as an argument to favor a somewhat smaller
value for $F/D$.
\m{
We should note that the relative sign of the $g_{N\Sigma K}$ and $g_{N\Lambda
K}$ couplings is often taken to be negative~\cite{Cot04}. It should be noted,
however, that this sign cannot be determined from a calculation as presented in
this work because it depends on an arbitrary sign in the definition of the
$\Lambda$-baryonic field. In weak-decay processes the sign of the vector v.s.\
axial vector can be determined, which only fixes the sign of the axial coupling
if the vector coupling is assumed to be positive~\cite{Cot04}.}

For the Baryon-Baryon-Vector (BBV, \tblref{summary-BBP}, lower plane)
sector the $SU(3)$ predictions are based on the assumptions of the extreme
quark model and vector meson universality~\cite{Swa89}, which requires $S=3F-D$
and $D=0$. With this choice for $SU(3)$ parameters the agreement
between $SU(3)$ predictions and values extracted from our calculation is
reasonable with the exception of the $\rho\Sigma\Sigma$ and $\rho\Sigma\Lambda$
couplings which are extracted to be much larger. A reason for this could be
that in the present calculation, in order to limit the number of free
parameters, the magnetic coupling of the $\rho$-meson, $\kappa_\rho$
was kept fixed for the different baryons.


\newcommand*{\XX}{\ensuremath{\Gamma'}}
\newcommand*{\YY}{\ensuremath{\Gamma}}

\begin{table}
  \caption{\tbllab{summary-BBP} \tbllab{summary-BBV} Baryon-meson summary table,
    defining
    $\YY(X)= (\chi X + i\gamma_\mu\partial^\mu X /2M)/(\chi+1) \gamF$
    and
    $\XX(X)= \gamma_\mu X^\mu +\frac{\kappa_X}{2M}(\sigma_{\mu\nu} \partial^\nu X^\mu )$.}
  \begin{ruledtabular}
  \begin{tabular}{C|C|d|d}
    \La   &  SU(3)  &\mc{$g_{SU(3)}$}&
      \mcc{$g_\text{model}$} \\
    \hline
    BBP \\
    \hline
      i g_{NN\pi} \bar N
       \YY(\dop{\vec\pi}{\vec\tau}) N
    & (D+F)/\sqrt2          &  13.47        & 13.47  \\ 

      i g_{NN\eta} \bar N
       \YY(\eta) N
    &(2S+3F-D)/3\sqrt2&   5.6         &  3.0   \\ 

      i g_{N\Lambda\K}
      \bar\Lambda
       \YY(\bar\K) N
    & (D+3F)/\sqrt6         &  13.3         & 12     \\ 

      i g_{N\Sigma\K}
      \bar\Sigma_i
       \YY(\bar\K \tau_i) N
    & (D-F)/\sqrt2          &   3.9         &   8.6  \\ 
    \hline
    BBV \\
    \hline
     -g_{NN\rho} \bar N 
      \XX(\dop{\vec\rho}{\vec\tau})
       N
    & (D+F)/\sqrt2        &    2.2          &   2.2  \\ 

     -g_{NN\omega} \bar N 
      \XX(\omega)
       N
    &(2S+3F-D)/3\sqrt2    &    6.6          &    8   \\ 

     -g_{NN\phi} \bar N 
      \XX(\phi)
       N
    & (3F-D-S)/3          &    0            &     0   \\ 

     -g_{N\Lambda K^*} \bar\Lambda 
      \XX(\bar \K^*)
       N
    & (D+3F)/\sqrt6       &    3.8          &   1.7  \\ 

     -g_{N\Sigma K^*} \bar\Sigma_i 
      \XX(\bar \K^*)
       \tau_i N
    & (D-F)/\sqrt2        &   -2.2          &  0     \\ 

     -g_{\Sigma\Sigma\rho} \eps_{ijk} \bar\Sigma_i 
      \XX(\rho_j)
       \Sigma_k
    & F\sqrt2          &       4.4          &  10    \\ 

     -g_{\Sigma\Lambda\rho} \bar\Sigma_i 
      \XX(\rho_i)
       \Lambda
    & -D\sqrt{2/3}     &       0            & -10    \\ 
  \end{tabular}
  \end{ruledtabular}
\end{table}


The values for the baryon magnetic moments as used in the calculation are
summarized in \tblref{summary-BBA} and are taken to agree with those quoted
by the Particle Data Group~\cite{PDG}. The value for $\kappa_{\Sigma^0}$ is
taken from the quark model prediction~\cite{Perkins}.
The $SU(3)$ parameters used are $\kappa_F=0.83$ and $\kappa_D=2.86$.
Note that the sign of the transition moment, $\kappa_{\Lambda\Sigma}$, is
chosen to agree with the $SU(3)$ prediction.


\begin{table}
  \caption{\tbllab{summary-BBA} Baryon-photon summary table.}
  \begin{ruledtabular}
  \begin{tabular}{C|C|d|d}
    \text{Vertex}     &  SU(3)   &\mc{$g_{SU(3)}$}&\mcc{$g_\text{model}$} \\
    \hline
    \kappa_p          & \kappa_F+\kappa_D/3   & 1.783 & 1.79  \\
    \kappa_n          & -2\kappa_D/3          &-1.907 &-1.91  \\
    \kappa_{\Sigma^+} & \kappa_F+\kappa_D/3   & 1.783 & 1.45  \\
    \kappa_{\Sigma^-} & -\kappa_F+\kappa_D/3  & 0.123 &-0.16  \\
    \kappa_{\Sigma^0} & \kappa_D/3            & 0.953 & 0.79  \\
    \kappa_{\Lambda}  & -\kappa_D/3           &-0.953 &-0.613 \\
    \kappa_{\Lambda\Sigma}& -\kappa_D/\sqrt3  &-1.651 &-1.61  \\
  \end{tabular}
  \end{ruledtabular}
\end{table}


\tblref{summary-VPP} summarizes the couplings in the meson sector, in
particular those for the Vector-Pseudoscalar-Pseudoscalar (VPP, upper plane)
and Vector-Pseudoscalar-Photon (VPA, lower plane) interaction Lagrangians.
For these sectors there are experimental data for meson decay widths~\cite{PDG},
which allows for the extraction of absolute magnitudes of coupling constants. For
the $SU(3)$ parameters we have chosen to adapt the predictions of the extreme quark
model, which dictates $G_{18}=\sqrt2 G_{88}$ for VPA and $G_{81}=0$ for the VPP
sectors of the model. As can be seen from the magnitude of the $\phi\pi\gamma$
coupling constant, this does not hold exactly but to a good extent. The
parameters in our calculation were fixed to the values extracted from
decay data where available, or to the $SU(3)$ predictions otherwise.
The same holds for the $\pi\gamma\gamma$ and $\eta\gamma\gamma$ coupling constants.


\begin{table*}
  \caption{\tbllab{summary-VPP} \tbllab{summary-VPA} Meson summary table,
  defining $\eps(ABCD)=(\eps_{\mu\nu\rho\sigma}(A^\rho B^\mu) (C^\sigma D^\nu))$}
  \begin{ruledtabular}
  \begin{tabular}{C|C|d|R|d}
         \La   &  SU(3)  &\mc{$g_{SU(3)}$}&
      \mc{$|g_\text{decay}|$}&\mcc{$g_\text{model}$} \\ 
    \hline
    VPP \\
    \hline
     -i g_{\phi\K\K} \bar \K \lrpart^\mu \K \phi_{\mu}
    &-G_{88}               & -4.5          &
            \X{4.48\pm0.07, K^\pm \\
               4.60\pm0.08, K^0 }                           &  -4.5  \\ 

     -i g_{\eta\K\K^*} \K \lrpart^\mu \eta \bar \K^*_\mu
    & G_{88}/\sqrt2-G_{81}\sqrt2& 3.2       &               &  -3.2  \\ 

     -i g_{\pi\K\K^*} \bar \K \lrpart^\mu \dop{\vec\pi}{\vec\tau} \K^*_\mu
    & G_{88}/\sqrt2         &  3.2          & 3.26\pm0.03   &  -3.26 \\ 

     -i g_{\rho\K\K} \bar\K \lrpart^\mu \tau_i \K \rho_{i\mu}
    & -G_{88}/\sqrt2        & -3.2          &               &    -3  \\ 

     -i g_{\rho\pi\eta} \dop{(\eta \lrpart^\mu \vec\pi)}{\vec\rho_\mu}
    & -G_{81}\sqrt2         &  0            & <2.7          &     0  \\ 

     -i g_{\rho\pi\pi} \eps_{ijk}
      {\vec\rho_{i\mu}}{(\vec\pi_j \lrpart^\mu \vec\pi_k)} /2
    & G_{88}\sqrt2          &  6.36         & 6.00\pm0.06   &   6.0  \\ 
    \hline
    VPA \\
    \hline
      \frac{g_{\rho\gamma\pi}}{m_\pi} \dop{\vec\pi^0}{
       \eps(\partial A \partial \vec\rho)}
    & G_{88}/3              &      -0.11    &0.12\pm0.02    &-0.12  \\ 

      \frac{g_{\rho\gamma\pi}}{m_\pi} \dop{\vec\pi^\pm}{
       \eps(\partial A \partial \vec\rho)}
    & G_{88}/3              &      -0.11    &0.101\pm0.006  &-0.10  \\ 

      \frac{g_{\omega\gamma\pi}}{m_\pi} \pi^0
       \eps(\partial A \partial \omega)
    &(G_{88}+G_{18}\sqrt2)/3&      -0.33    &0.322\pm0.007  & 0.32   \\ 

      \frac{g_{\phi\gamma\pi}}{m_\pi} \pi^0
       \eps(\partial A \partial \phi)
    &(G_{88}\sqrt2-G_{18})/3&       0.0     &0.018\pm0.001  & 0.018 \\ 

      \frac{g_{\phi\gamma\eta}}{m_\pi} \eta
       \eps(\partial A \partial \phi)
    &(2G_{81}-G_{18}-G_{88}\sqrt2)/9&  0.1  &0.096\pm0.002&0.096 \\ 

      \frac{g_{\K^*\K\gamma}}{m_\pi} \bar\K^0
       \eps(\partial A \partial \K^*)
    & -2G_{88}/3            &       0.22    &0.177\pm0.009  &-0.177  \\ 

      \frac{g_{\K^*\K\gamma}}{m_\pi} \bar\K^\pm
       \eps(\partial A \partial \K^*)
    & G_{88}/3              &      -0.11    &0.117\pm0.007  & 0.117  \\ 
  \end{tabular}
  \end{ruledtabular}
\end{table*}


\newcommand*{\elp}[4]{\displaystyle\frac{#1}{#2}+\frac{#3}{#4}}
\newcommand*{\elm}[4]{\displaystyle\frac{#1}{#2}-\frac{#3}{#4}}

\section{$SU(3)$ notation}
\label{app:su3}

The notation used to derive the $SU(3)$ model couplings is summarized.
Our notation agrees with that of ref.~\cite{Tit99}, and a more
detailed review of $SU(3)$ can be found in ref.~\cite{Swa63} or ref.~\cite{Mos89}.

The definition for pseudo-scalar singlet is
\begin{equation}
  P_1 = \diag(1,1,1)\eta_1 \,,
\end{equation}
and that for the octet
\begin{equation}
  P_8=\begin{pmatrix}
    \elp{\pi^0}{\sqrt{2}}{\eta_8}{\sqrt{6}}  &   \pi^+    &    K^+\\
    \pi^-     & -\elp{\pi^0}{\sqrt{2}}{\eta_8}{\sqrt{6}} &    K^0\\
    K^-       & \bar{K}^0                         & -2\eta_8/\sqrt{6}
  \end{pmatrix} \,,
\end{equation}
where in the quark model $\pi^0=(u\bar{u} - d\bar{d})/\sqrt{2}$,
$\eta_1=(u\bar{u} + d\bar{d} + s\bar{s})/\sqrt{3}$, and
$\eta_8=(u\bar{u} + d\bar{d} - 2s\bar{s})/\sqrt{6}$.
The photon field couples to the charge, which in SU(3) language is
\begin{equation}
  \hat{Q}=\diag(2/3,-1/3,-1/3) \;.
\end{equation}
The baryon octet is given by
\begin{equation}
  \begin{split}
  B_8&=\begin{pmatrix}
    \elm{\Sigma^0}{\sqrt{2}}{\Lambda}{\sqrt{6}}  &   \Sigma^+   &   p\\
    \Sigma^-     & -\elm{\Sigma^0}{\sqrt{2}}{\Lambda}{\sqrt{6}} &   n\\
    \Xi^-                                &   \Xi^0      & +2\Lambda/\sqrt{6}
  \end{pmatrix} \\
  &=\begin{pmatrix}
    uds & uus & uud\\
    dds & uds & udd\\
    dss & uss & uds
  \end{pmatrix} \,.
  \end{split}
\end{equation}

The Lagrangian for the different $SU(3)$ sectors read
\begin{align}
  \La_{VPP}&=G_{88} \Tr \Big[
    V_8^\mu (\dop{P_8}{\partial_\mu P_8} - \dop{\partial_\mu P_8}{P_8})
    \Big] \\
   &\quad+G_{81}\sqrt{3} \Tr \Big[
    V_8^\mu (\dop{P_8}{\lrpart_\mu p_1})
    \Big] \notag\\
  \La_{VPA}&=i \eps_{\mu\nu\alpha\beta} k_V^\alpha (\partial^\mu A^\nu) \Biggl[
    G_{81}\sqrt{\frac23} \Tr \Big[ \hat{Q} V_8^\beta P_1 \Big] \\
   &+G_{18}\sqrt{\frac23} \Tr \Big[ \hat{Q} V_1^\beta P_8 \Big]
    +G_{88} \Tr\Big[ \hat{Q} \big\{ V_8^\beta, P_8 \big\}_+ \Big]
    \Biggr] \notag\\
  \La_{BBP}&=D \Tr \Big[P_8 \{B_8,\bar{B}_8 \}_+ \Big] \\
     &\quad+F \Tr \Big[P_8 [B_8,\bar{B}_8 ]_-   \Big]
     +\frac{S}{\sqrt{3}} \Tr \Big[P_1 \bar{B}_8 B_8 \Big] \notag
\end{align}

The physical particles are related to the pure octet and singlet states as
\begin{equation}
  \begin{split}
    \eta =&\eta_8 \cos\theta_P + \eta_1 \sin\theta_P \\
          &\approx (\eta_8 + \sqrt{2}\eta_1)/\sqrt{3}
          = (u\bar{u} +d\bar{d})/\sqrt{2}\\
    \eta'=&\eta_8 \sin\theta_P - \eta_1 \cos\theta_P \\
          &\approx (\sqrt{2}\eta_8 - \eta_1)/\sqrt{3}
          = -s\bar{s}
  \end{split}
\end{equation}
for $\theta_P=60^\circ - \Delta \theta_P$. For vector particles,
$V$ (scalar, $S$) similar definitions apply replacing only
$\pi \to \rho (a_0)$, $\eta \to \omega (\sigma)$,
$\eta' \to \Phi (f_0)$, $K \to K^* (K^*_0)$.

\end{document}